\documentclass[a4paper,11pt]{article}
\pdfoutput=1 

\usepackage{jcappub} 

\usepackage[T1]{fontenc} 

\usepackage[sort&compress]{natbib}
\usepackage[utf8]{inputenc}
\usepackage{multirow}
\usepackage[english]{babel}
\usepackage{hyperref}
\usepackage{amsmath}
\usepackage{amssymb}
\usepackage{epsfig}
\usepackage{ccaption}
\usepackage{lscape}
\usepackage{footnotebackref}
\usepackage{longtable}
\usepackage{float}
\usepackage{graphics,psfrag,rotating}
\usepackage{graphicx}
\usepackage{dcolumn}
\usepackage{bm}
\usepackage{array,multirow}
\usepackage{xspace}
\usepackage{xcolor}
\usepackage{placeins}
\usepackage{afterpage}
\usepackage{etoolbox}

\newcommand{\MLR}{\Lambda_{\mathcal{M}}}

\begin{document}

\title{Bayesian reconstruction of the Milky Way dark matter distribution}


\author[a,b,e]{E.V.~Karukes},
\author[b]{M.~Benito},
\author[a,b]{F.~Iocco},
\author[c,d]{R.~Trotta},
\author[c]{A.~Geringer-Sameth},


\affiliation[a]{ICTP-SAIFR \& IFT-UNESP, R. Dr. Bento Teobaldo Ferraz 271, S\~ao Paulo, Brazil}
\affiliation[b]{IFT-UNESP, R. Dr. Bento Teobaldo Ferraz 271, S\~ao Paulo, Brazil}
\affiliation[c]{Physics Department, Astrophysics Group, Imperial Centre for Inference and Cosmology, Blackett Laboratory, Imperial College London, Prince Consort Rd, London SW7 2AZ}
\affiliation[d]{Data Science Institute, William Penney Laboratory, Imperial College London, London SW7 2AZ}
\affiliation[e]{Astrocent, Nicolaus Copernicus Astronomical Center Polish Academy of Sciences, ul. Bartycka 18, 00-716 Warsaw, Poland}

\emailAdd{ekarukes@camk.edu.pl}
\emailAdd{mariabenitocst@gmail.com}
\emailAdd{fabio.iocco.astro@gmail.com}
\emailAdd{r.trotta@imperial.ac.uk}
\emailAdd{a.geringer-sameth@imperial.ac.uk}

\newcommand{\FI}[1]{{\color{blue}[{\bf FI:} #1]}}
\newcommand{\EK}[1]{{\color{red}[{\bf EK:} #1]}}
\newcommand{\MB}[1]{{\color{orange}[{\bf MB:} #1]}}
\newcommand{\RT}[1]{{\color{magenta}[{\bf RT:} #1]}}
\newcommand{\com}[1]{{\color{green}[{\bf comment:} #1]}}
\newcommand{\rt}[1]{\RT{#1}}
\newcommand{\AGS}[1]{{\color{cyan}[{\bf AGS:} #1]}}
\newcommand{\sigmaint}{\sigma_\text{int}}

\abstract{We develop a novel Bayesian methodology aimed at reliably and precisely inferring the distribution of dark matter within the Milky Way using rotation curve data. We identify a subset of the available rotation curve tracers that are mutually consistent with each other, thus eliminating data sets that might suffer from systematic bias. We investigate different models for the mass distribution of the luminous (baryonic) component that bracket the range of likely morphologies. We demonstrate the statistical performance of our method on simulated data in terms of coverage, fractional distance, and mean squared error. Applying it to Milky Way data we measure the local dark matter density at the solar circle $\rho_0$ to be $\rho_0 = 0.43\pm 0.02(\rm{stat})\pm0.01(\rm{sys})$~GeV/cm$^3$, with an accuracy $\sim$~6\%. This result is robust to the assumed baryonic morphology. The scale radius and inner slope of the dark matter profile are degenerate and cannot be individually determined with high accuracy. We show that these results are robust to several possible residual systematic errors in the rotation curve data.}
 
\maketitle
\flushbottom

\section{Introduction}
\label{sec:intro}

Dark matter is a crucial ingredient for the formation of structures in our universe. Without dark matter it would be impossible for small density perturbations to grow into the huge potential wells that host the galaxies we observe today. Dark matter thus constitutes the invisible gravitational backbone of our universe. It cannot, however, be accommodated within the Standard Model of particle physics and large experimental and theoretical efforts are underway to identify its nature. At the same time astronomical observations of increasing refinement are characterizing the distribution of dark matter within the visible galaxies.
On one hand, such observations may be useful to characterize ancillary quantities required to interpret dark matter detection experiments. On the other, the distribution of dark matter within galaxies of different sizes is a prediction of the $\Lambda$CDM model, and measuring these distributions through astrophysical analyses provide an important test of consistency of the cosmological framework.

Different methods are adopted in the literature to probe the dark matter distributions in galaxies and clusters of diverse sizes, ranging from gravitational lensing for clusters \cite{2012ApJ...755...56U,2013ApJ...765...25N,2015ApJ...814...26N}, Jeans analysis of the motions of stars for dwarf and elliptical galaxies \cite{Lokas:2001mf,2013pss5.book.1039W,Battaglia:2013wqa}, and the motion of tracer objects in circular orbits in disk galaxies \cite{1980ApJ...238..471R,2006ApJS..165..461K,Karukes:2015fma}. The latter is often referred to as the Rotation Curve (RC) method, and relies on the assumption that the stellar disk of a galaxy is rotation supported, and that the tracers are in circular motion around the gravitational center of the galaxy. These assumptions typically hold in disk galaxies over a large range of masses including our very own, the Milky Way.

The dark matter distribution (often referred to as the ``density profile'') is inferred by fitting the observed velocity distribution (as a function of the galactocentric distance) with the one expected from the observed, luminous matter plus the ``{\it missing component}'' \cite{ede6256c6323491dbfc5ffbbc4efe413}.

The values (and uncertainties) resulting from this procedure are usually adopted, and a dark matter density profile for the Galaxy at hand assumed from this. Studies in the literature \cite{Sofue:2000jx, Catena:2009mf,Iocco:2011jz, Nesti:2013uwa,Silverwood:2015hxa} usually assume spherical distribution for dark matter, and one (or at most few) possibilities for the underlying visible component.

In this paper we revisit this well known method. First we propose a new Bayesian procedure to characterize the Milky Way's dark matter density profile based on the current state-of-the-art for both the observed rotation curve and the luminous (hereafter, baryonic) component of the Milky Way. In this context, we also study the possible incompatibility of different data sets using a Bayesian procedure in order to rule out the possibility of systematic discrepancies between data sets.

Second, we study the {\it accuracy} of such a method: we propose for the first time the validation of the RC method in the Milky Way by generating mock rotation curves constructed on the basis of a known, controlled dark matter profile, and holding the same statistical properties as the observed ones. We test their reconstruction through the RC method against the known truth. The two separate procedures that we have developed and present here can be easily applied to new incoming data sets for the Milky Way, as well as to other galaxies.

The paper is structured as follows:
in Section~\ref{sec:astro} we present our astrophysical setup, namely the set of observations adopted for the rotation curve and for the luminous (baryonic) component of the Milky Way; in Section~\ref{sec:method} we present our methodology for reconstruction and describe our statistical and numerical tools. Section~\ref{sec:results} contains an extended discussion on our main findings, and we summarize our conclusions in Section~\ref{sec:conclusions}. 
In Section~\ref{sec:real data} we directly compare the results of our procedure for three  extreme  baryonic morphologies, while in Appendix \ref{App:morphologies} we show the results for the whole set of different baryonic morphologies collected in the literature and presented in \cite{Iocco:2015xga}.

\section{Astrophysical setup}
\label{sec:astro}

The rotation curve (RC) of a galaxy provides important constraints on its mass distribution, including its dark matter content (see e.g. \cite{vanAlbada:1984js,1979A&A....79..281B, Sofue:2000jx}). Paradoxically, the RC of the Milky Way is much harder to measure than those of external galaxies. This is due to our interior position that complicates some measurements, such as the extended RC of the gas in its disk. Hence, in order to measure the RC of both the inner and outer regions one can use two different classes of tracers that we summarize below.

In what follows we adopt the following fixed Galactic parameters $R_0=8.34$ kpc,  $V_0=239.89\;\rm{km/s}$, ($U_\odot$, $V_\odot$, $W_\odot$)=(7.01, 12.20, 4.95) $\rm{km/s}$. These are, respectively, the galactocentric distance to the Sun, the circular velocity at the Sun's position, and the local standard of rest. Given these fixed parameters, we explore several important sources of uncertainty, including systematic errors in the rotation curve data, the dependency on the morphological models we assume for the disk component, and the observational uncertainties on the normalizations of the baryonic components.

It is to be noticed that the uncertainty existing in the literature about the Galactic parameters
does not impact the general conclusions on the presence of dark matter even within the solar circle (see Supplementary Information of~\cite{Iocco:2015xga} for details). However, such uncertainties can affect the actual determination of the dark matter distribution (see e.g.~\cite{2015JCAP...12..001P}), and therefore they must be taken into account by dedicated analysis, also in light of its effects on searches for the very nature of the dark matter \cite{2017JCAP...02..007B, Belyaev:2018pqr}. A quantitative investigation of the effect of Galactic parameter uncertainties will be addressed in a future work~ \cite{Benito:2019ngh}.

\subsection{The observed rotation curve
\label{sec:observed RC}}

\begin{figure}
 \begin{centering}
\includegraphics[angle=0,height=9.truecm,width=14.truecm]{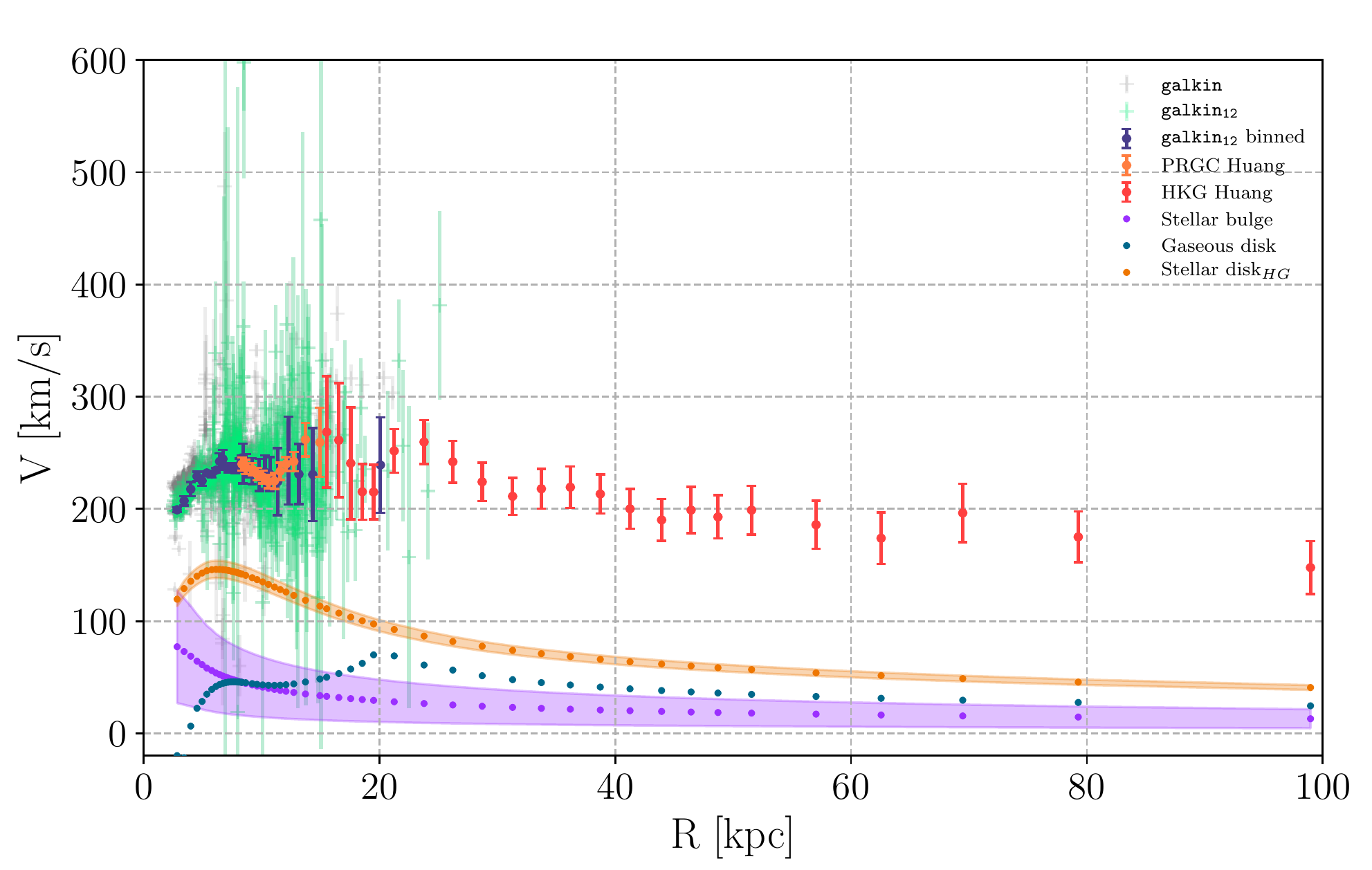}
\caption{RCs measurement from \cite{2016MNRAS.463.2623H} and \cite{Pato:2017yai}. The gray errorbars represent the full \texttt{galkin} data compilation (25 data sets), which is reduced to the  \texttt{galkin}$_{12}$ compilation, comprising 12 data sets, by demanding statistical consistency between the constituent data sets. The \texttt{galkin}$_{12}$ compilation is plotted in light green and the corresponding binned RC is shown in purple. Orange and purple shadowed areas denote the 1$\sigma$ observational uncertainties of the stellar disk and bulge for our reference morphology, respectively (the uncertainties are propagated along the radii only for the visualization purpose). The uncertainties reflect the measured errors on  the values of optical depth and stellar surface mass density. Blue points represent distribution of the gas.}
\label{fig:RCs1}	 
\end{centering}
\end{figure}

For the innermost region of the Galaxy (galactocentric radii in the range $2.5 - 22$ kpc) we adopt the \texttt{galkin} data compilation, first presented as a whole in \cite{Iocco:2015xga}.
This compilation includes kinematics of gas, masers, and stars for a total of 2780 tracers in the disk, separated into 25 individual datasets, collected from almost four decades of literature.
We address the reader to the public release of this data (and of the online tool to manage them) \cite{Pato:2017yai} for further details on the individual datasets, their nature, and original literature sources.

A close inspection of the data (plotted as gray data points in Fig.~\ref{fig:RCs1}) shows that, due to their heterogeneous nature, some individual data sets show marked discrepancies with others. For instance, in the region around $R=10~\rm kpc$ we observe data points that are clearly discrepant (by many standard deviations) from the bulk of the observations. Such data points might suffer from distinctive systematic effects, or other data-collection artifacts, and it would be unwise to simply bin all the data together lest the outliers exert a disproportionate (and incorrect) influence on the bin average. In particular, most data sets that we have excluded are based on observations of such tracers as HI regions and CO emission associated with HII regions. The offsets between different data sets based on similar type of tracers might arise due to several reasons, as for instance: differences in the method used to estimate the maximum line-of-sight velocities (see discussion in \cite{McClureGriffiths:2007ts}), differences in the Galactic longitudes at which data were taken (the long-known asymmetry between the southern and the northern rotation curves, see e.g. \cite{1991ApJ...370..205B,2008ApJ...679.1288L}).
Therefore it is necessary to invoke a statistical procedure to decide which subset of the 25 data sets are mutually compatible. The aim of the procedure is to weed out highly discrepant data sets, whose inclusion would systematically bias the bin value. To this end, we use a method that uses the Bayesian evidence (or model likelihood) to determine the degree of overlap in parameter space between parameter constraints (in a given parametric model) obtained using two different data sets. The idea is that if the data sets are compatible, constraints obtained by using them jointly will overlap significantly with those obtained when using each data set individually (see e.g. \cite{Hobson:2002zf,Trotta+2011}). The details on the procedure are given in Section~\ref{sec:bayesian_evidence} below. 

After determining the subset of data within \texttt{galkin} that is mutually compatible via the above procedure, we carry out error-weighted binning of the data in radius. This is necessary in order to use a binned likelihood approach across the entire range of radial distances. We require an approximately equal number of data points per bin in order to obtain well-balanced bins errors. The number of bins is chosen such that the size of it is approximately equal to the bin size of \cite{2016MNRAS.463.2623H}. This yields to 30 bins in total.  The mean value in each bin is calculated using the weighted mean, and the corresponding standard deviation by using the weighted standard deviation:
\begin{equation}\label{eqn:binning}
\bar V= \frac{\sum^{N_{bin}}_{i=1} V_i/\sigma^2_{i}}{\sum^{N_{bin}}_{i=1}1/\sigma^2_{i}}\hspace{0.5cm} \mathrm{and}\\
\hspace{0.5cm}\sigma_{\bar V}^2 =
\frac{\sum ^{N_{bin}}_{i=1}(\bar V - V_i)^2/\sigma^2_{i}}{\sum^{N_{bin}}_{i=1}1/\sigma^2_{i}}\\
\end{equation}
where $\bar V$ is the weighted mean of the observed circular velocity,  $\sigma_{i}$ is the uncertainty in the circular velocity measurement $V_i$, and $N_{bin}$ is the number of data points of the $i^{\rm th}$ bin. We have verified that this procedure leads to a scatter within each bin that closely follows a Gaussian distribution. This is a necessary condition for our likelihood function (defined later on) to be a good representation of the data in each bin.

 For the intermediate region of the Milky Way we use a data set obtained from the data compilation of Huang~et~al.~\cite{2016MNRAS.463.2623H}. The authors use $\sim$~16,000 primary red clump giants (PRCGs) to derive a binned determination of the RC of the outer stellar disk (galactocentric distances in the range $8.34 - 14.95$~kpc) (see Fig. \ref{fig:RCs1}). 

For the outer Milky Way (with radius in the range $\sim 15$ to $\sim$~100~kpc) we use the binned RC measurements obtained by Huang~et~al.~\cite{2016MNRAS.463.2623H} from $\sim$5,700 so-called halo K giant (HKG) stars. These stars, differently from the tracers in the disk, do not follow circular orbits. Therefore, in order to derive the RC one needs to apply the Jeans analysis (see \cite{2016MNRAS.463.2623H} for more details). We plot the resulting RC measurement on Fig.~\ref{fig:RCs1}.

\subsubsection{Compatibility of data sets via the Bayesian evidence}  
\label{sec:bayesian_evidence}
  
 Our method to select mutually compatible data sets uses a Bayesian model comparison framework. For further details about the methodology, see \cite{Hobson:2002zf,Marshall:2004zd,Trotta+2011,Feroz:2008wr,Feroz:2009dv,PhysRevD.93.103507,Abbott:2017wau,Handley:2019wlz}. Given a data set $\rm d_1$, we consider whether a second data set $\rm d_2$ is compatible, assuming that the two data sets are statistically independent. We perform a model comparison and evaluate the model likelihood ratio $\MLR$, defined as
\begin{equation}\label{eqn:evidence}
\MLR = \frac{p(\rm{d_1},\rm{d_2}|\mathcal{M})}{p(\rm{d_1}|\mathcal{M})p(\rm{d_2}|\mathcal{M})},
\end{equation}
\noindent
 where $p(\rm{d}|\mathcal{M})$ is the Bayesian evidence (or model likelihood) for data $\rm{d}$ given model $\mathcal{M}$. In the above, $\rm{d}_1$, $\rm{d}_2$ are the two data sets being tested for compatibility; model $\mathcal{M}$ has parameters $\Theta$ (as described below in Section \ref{sec:method}), and in the numerator $p(\rm{d_1},\rm{d_2}|\mathcal{M})$ is the probability of obtaining both data sets if they both come from the same underlying parameters. The denominator is the probability of obtaining both data set when each of them is separately described by its own set of parameters, $\Theta$. The model likelihood ratio $\MLR$ thus represents the change in odds ratio induced by the data when comparing a model where both data sets are described by a single set of parameters and are thus compatible, to a disjoint model  where each of the data sets is described by its own set of parameters.  If $\MLR>1$, then the two data sets are deemed to be compatible (as their induced parameter space constraints overlap), and data set $\rm{d}_2$ is added to the list of compatible data sets. If $\MLR<1$ data set $\rm{d}_2$ is deemed incompatible with the data sets previously accepted and it is discarded.  The details on the model we use and the form of the likelihood are given in Section~\ref{sec:method}.

The procedure begins by selecting an initial data set from \texttt{galkin}. We choose the measurements of Malhotra~et~al.~\cite{Malhotra:1994qj} since by inspection it appears smooth with no outliers and covers the inner $\sim$~8~kpc of the Milky Way RC. Then, we consider additional data sets one by one and use Eq.~\ref{eqn:Bayes} to decide whether to include them. We end up with 12 compatible data sets out of 25, which we denote \texttt{galkin}$_{12}$. The latter contains the following data sets: gas kinematics \cite{Malhotra:1994qj,1985AJ.....90..254K,1985ApJ...295..422C}, star kinematics \cite{2007A&A...473..143D,2013Ap.....56...68B,1994A&A...285..415P,1997A&A...318..416P} and masers \cite{2014ApJ...783..130R,2012PASJ...64..136H,2013ApJ...769...15X,Stepanishchev2011,2013AstL...39..809B}. However, it is worth noticing that the order in which the data sets are tested can be important. Indeed, if one starts from (or includes at the first steps) a data set that is incompatible with the rest (or with most of the others), then the final number of the compatible data sets is smaller\footnote{Notice that we did not perform this test for all the possible combinations of the 25 data sets since we deal with time-consuming computations.} than 12. We have checked that among the 12 selected data sets the order of inclusion is not important, demonstrating the robustness of the procedure. We then bin the selected subsample in the manner described in Subsection~\ref{sec:observed RC}. The resulting binned RC is shown in Fig.~\ref{fig:RCs1}. We check to see whether the derived RC is compatible with the RC measurements we adopt for the outer Milky Way~\cite{2016MNRAS.463.2623H}. However, notice that the outermost data of \cite{2016MNRAS.463.2623H} separately are not compatible with its inner part as well as with our binned subsample \texttt{galkin}$_{12}$.  Therefore, we perform an error bar re-scaling analysis that has the ability to correct for any systematic underestimation of the measurement uncertainties that might be present in our adopted data sets (see Section~\ref{sec:real data}).

The model comparison procedure weeds out a sizable fraction of data from the {\tt galkin} compilation, and the results of the analysis with the {\it whole} {\tt galkin} compilation versus the ``{\it trimmed}'' one differ by a sizable factor of approximately 50 \% in $\rho_0$.
We therefore use the ``{\it trimmed}'' {\tt galkin}$_{12}$ compilation for our analysis even if the power of the Huang~et~al.~\cite{2016MNRAS.463.2623H} data -once included- drives the final results thus making the ``full {\tt galkin} + Huang'' virtually indistinguishable from the ``{\tt galkin}$_{12}$ + Huang''.

 We also note that the Huang~et~al.~\cite{2016MNRAS.463.2623H} data have a dip in circular velocity between 10-20~kpc that is also noticeable in the \texttt{galkin} compilation. The origin of the dip is currently unclear. The authors of \cite{2016MNRAS.463.2623H} state that this dip can be either an artifact due to high measurement uncertainties or it might reflect some actual feature of the kinematic tracers. Therefore, in order to check whether this dip in the data has any impact on the model fitting results, we perform a test where we exclude from our analysis  the data between 10-20~kpc (see Section~\ref{sec:results} for the results).

We have also carried out a consistency check by adopting different --- and extreme --- baryonic morphologies: 
for the central, heaviest, and lightest disks (namely those that provide the central, highest, and lowest RC) we have performed the procedure to select mutually compatible {\tt galkin} data sets.
The compatible data sets resulting from the procedure are always the same 12, and the inferred value of the local density $\rho_0$ remains constant, its precision
 staying virtually unchanged (the same does not apply to the anyway poorly constrained scale radius $r_s $ and density profile inner slope $\gamma$, see Section \ref{sec:results} and Appendix \ref{App:morphologies} for details), thus validating the procedure as insensitive to the various baryon morphologies.

\subsubsection{The visible component (baryonic morphology) \label{sec:baryons}}

In order to constrain the Milky Way's dark matter content and its distribution, we need to include a model of the visible (baryonic) component of the Milky Way.
Despite the improvements in the amount and quality of observations, controversies about the actual distribution of visible matter in the Milky Way remain. The uncertainties on the distribution of baryonic components do in turn impact the determination of the dark matter distribution, as addressed in detail by Iocco et. al.~\cite{Iocco:2015xga}. 

Following the approach of~\cite{Iocco:2015xga,Pato:2017yai,Iocco:2016itg}, we account for the stellar bulge, stellar disk(s), and gaseous disk using a fully observation-driven approach. The mass density of each one of the components is described by a three-dimensional function inferred from the observation of the stellar and gas target populations. Different functional shapes are available in the literature for both the disk(s) and the bulge component, with the latter presenting the highest degree of variation (it is to be noted that the differences are most important at small galactocentric distances).
The diversity of observationally inferred morphologies of the baryonic component, available to date in the literature, must necessarily be accounted for as it impacts final conclusions on the matter distribution, as shown in previous work, e.g.~\cite{2015JCAP...12..001P}. We therefore collect a vast array of observationally inferred morphologies, separately for the bulge and the disk, and describe their treatment to our purposes in the following. 

It is noteworthy that whilst all disks (thin, and thick) have necessarily cylindrical symmetry, the bulge morphologies show no degree of symmetry, presenting bar(s) at different angles (we devote to the bulges a fully consistent 3--dimensional treatment). The array of disk(/bulge) morphologies that we adopt here is taken from \cite{2015JCAP...12..001P}, where a summary of the different morphologies and of their normalization has been originally presented, and the interested reader may find appropriate citation to the original sources of observations for each of them.
As of the use of such large array of possibilities for the visible component, rather than opting for choosing a single choice of bulge and disk, which would necessarily force us into a strong assumption on the baryonic components' distributions, we follow the wake of previous analysis \cite{2015JCAP...12..001P,Iocco:2016itg,2017JCAP...02..007B,Benito:2019ngh}, and adopt each possible combination of bulge and disk morphologies independently, thus constructing a catalogue of alternative ``{\tt baryonic morphologies}'' out of all the possible individual combinations of the two (plus the gas disk).  By treating each baryonic combination as individual and independent we are able to bracket the systematic uncertainties that might alter the inferred dark matter density distribution~\cite{Iocco:2015xga,2015JCAP...12..001P}. 
Along these lines, we first adopt one single morphology to test our methodology and then test the generality of our conclusions on the methodology for the most extreme configurations allowed in the literature, namely those producing the ``heaviest'' and ``lightest'' rotation curves, see Section~\ref{sec:real data}.

In what follows we introduce the reference baryonic morphology that we adopt throughout this study. Additionally, we present posterior results when adopting two other baryonic morphologies that are expected to bracket the possibilities for the disk and bulge.
Finally, in Appendix~\ref{App:morphologies} we introduce the details and we show the model fitting results for all baryonic morphologies presented in \cite{Iocco:2015xga}, thus bracketing the systematic uncertainty due to the observational ignorance on the morphology of the visible component of the Milky Way.

{\bf Stellar bulge.} The bulge dominates the inner 2-3~kpc of the Milky Way and presents a triaxial shape with
a bar extending at positive Galactic longitudes. This
general picture is consistently painted by different observations, but the morphological details are rather uncertain. Following~\citep{Dwek:1995xu} we implement their E2 triaxial mass density model of the bulge that is based on COBE photometric maps. This model and the further calculations of the corresponding circular velocity have been presented in detail elsewhere~\cite{Iocco:2011jz}. We follow the same type of calculation as in~\citep{Iocco:2011jz} and all the parameters of the E2 fitting formula except the total normalization are fixed according to~\citep{Stanek:1996qy}. The normalization of the mass density of the stars in the bulge, or more concretely the microlensing optical depth $\left<\tau \right>$  is kept as a free parameter (but constrained with data from MACHO \cite{Popowski:2004uv}). The microlensing optical depth towards the galactic bulge defines the total mass density of stars in the bulge (see for more details~\citep{Iocco:2011jz} on the relation of the microlensing optical depth and the bulge density normalization) and it is a measured quantity with an uncertainty of about 20\%.

{\bf Stellar disk.} It is well-known that the Milky Way stellar disk consists of two components: a thin and a thick disk that correspond to distinct stellar populations. In the present analysis we adopt a double exponential disk mass profile where, according to~\cite{Han:2003ws}, the thickness of the thin (thick) disk and the exponential disk length scale are $\rm{z_h^{thin}}=0.27$~kpc~$(\rm{z_h^{thick}}=0.44$~kpc)  and $R_d=2.9$~kpc, respectively\footnote{For consistency, we provide the values that have been recalculated adopting $R_0$ = 8.34~kpc}. Meanwhile, in our analysis, we keep free the local stellar surface mass density $\Sigma_*$. 

{\bf Gas.} The gas component is a sub-dominant component of the Milky Way. It takes the form of molecular, atomic and ionized hydrogen and its distribution along with its total mass density are relatively well-known.
We follow the observationally inferred morphology for the density distribution of the gaseous disk taken from~\cite{Ferriere:2007yq} for the inner 3~kpc and from~\cite{0004-637X-497-2-759} for the remaining range. We do not take into account the uncertainties on the density distribution profile of the gaseous disk. This is due to both its dynamically sub-dominant nature and its relatively well measured density distribution.

Finally, in Fig.~\ref{fig:RCs1} we plot circular velocities that correspond to the above-described baryonic components. The shaded regions in Fig.~\ref{fig:RCs1} illustrate the observational uncertainties on the normalization factors $\left<\tau\right>$ and $\Sigma_*$ (see Section \ref{sec:method} for details) in the inferred bulge and disk RCs.

\subsection{The dark matter component}\label{sec:DM}

As a model of the dark matter distribution within the Galaxy we adopt the spherical generalized Navarro, Frenk, and White (gNFW) profile~\cite{Zhao:1995cp,2001ApJ...555..504W}:

\begin{equation}\label{eqn:gNFW}
\rho_{gNFW}(r)=\rho_0\left(\frac{R_0}{r}\right)^{\gamma}\left(\frac{r_s+R_0}{r_s+r}\right)^{(3-\gamma)},
\end{equation}

\noindent where $r_s$ is the characteristic radius of the halo, $\rho_0$ is the dark matter density at Sun's location $R_0=8.34$~kpc, and $\gamma$ is the inner slope of the density profile\footnote{The value $\gamma=1$ corresponds to the standard NFW profile.}. The enclosed mass within a sphere of radius $R$ is

\begin{equation}\label{eqn:mass_sph}
M_{DM}(<r)=4 \pi \int_0^r  \rho_{gNFW}(r)\hspace{0.1cm}R^2\hspace{0.1cm}dR.
\end{equation}

\noindent
In the assumed spherical dark matter distribution the angular velocity for a circular orbit is

\begin{equation}\label{eqn:vel_dm}
\omega_{DM}^2=\frac{G M_{DM}(<r)}{r^3}.
\end{equation}

\section{Statistical framework \label{sec:method}}

The RC curve obtained in Sections~\ref{sec:observed RC}--\ref{sec:DM} reflects the total (baryonic + dark matter) kinematics of the Milky Way.
We now use this RC, in combination with information on the distribution of gas, stars and dark matter as described in Section~\ref{sec:astro}, to perform the global mass modeling and to gather insights on the  underlying dark matter distribution. To do so we fit a global model of the Galaxy that consists of four components:  stellar disk, gaseous disk, stellar bulge and dark matter halo,

\begin{equation}
\omega_{c}^2(\Theta, \Sigma_*, \langle\tau\rangle)=\omega_{disk}^2(\Sigma_*)+\omega_{gas}^2+\omega_{bulge}^2(\Sigma_*, \langle\tau\rangle)+\omega^2_{DM}(\Theta),
\label{eqn:vel_model}
\end{equation}
\noindent
where the first three baryonic components are presented in Section~\ref{sec:baryons} and the dark matter component (see Section~\ref{sec:DM}) depends on parameters $\Theta = (\rho_0, r_s, \gamma)$. Note that each term in Eq.~\ref{eqn:vel_model} is implicitly a function of galactocentric radius $r$ and that angular velocity $\omega_c$ is used instead of linear velocity $V_c\equiv r\omega_c$.

\subsection{Priors, likelihood, and posterior sampling \label{sec:MCMC}}

For a given morphology, our model has five free parameters: the scale radius of the dark matter halo $r_s$, the inner dark matter density slope $\gamma$, the local dark matter density $\rho_0$, the microlensing optical depth $\left<\tau\right>$, and the stellar surface density $\Sigma_*$. We explore the model parameter space using uniform priors in the following variables and ranges:

\begin{equation}\label{eqn:priors}
\begin{aligned}
    0<r_s/{\rm[kpc]}<40, \\
    0<\gamma<3, \\
    0<\rho_0/{\rm[GeV\hspace{0.2cm}cm^{-3}]}<1, \\
    19\times 10^6<\Sigma_*{\rm[M_{\odot}\hspace{0.2cm}kpc^2]}<57\times 10^6, \\
    0.1\times10^{-6}<\left<\tau\right><4.5\times 10^{-6}.
\end{aligned}
\end{equation}
\noindent
We use fairly wide priors, which encompass the support of the likelihood. The last two parameters ($\Sigma_*$ and $\left<\tau\right>$), are nuisance parameters which are each independently constrained by a Gaussian likelihood. For the mean and standard deviation of this likelihood we adopt the values of the stellar surface density at the Sun's position provided by~\cite{2013ApJ...779..115B}, i.e. $\Sigma_*^{obs}=(38\,\pm\, 4)\times10^6\,\rm{M_{\odot}/kpc^2}$, as well as the measurement of the microlensing optical depth provided by the MACHO collaboration in Popowski~et~al.~\cite{Popowski:2004uv}, namely $\langle\tau\rangle^{obs}=2.17^{+0.47}_{-0.38}\times 10^{-6}$. For simplicity, we symmetrize the error in the microlensing optical depth and adopt a standard deviation of $\sigma_{\langle\tau\rangle}=\pm0.47$ (which is conservative, as it uses the larger of the two errors).

The likelihood function takes the form

\begin{equation}
\begin{split}
P(d|\Theta, \Sigma_*, \langle\tau\rangle) & = 
\prod_{i=1}^m\left\{\frac{1}{\sqrt{2\pi}\sigma_{\bar{\omega},i}}\exp\left[-\frac{1}{2}\frac{\left(\omega_c(r_i, \Theta, \Sigma_*, \langle\tau\rangle)-\bar{\omega}_i\right)^2}{\sigma_{\bar{\omega},i}^2}\right]\right\} \\
&\times \frac{1}{\sqrt{2\pi}\sigma_{\langle\tau\rangle}}\exp{\left[-\frac{1}{2}\frac{\left(\langle\tau\rangle - \langle\tau\rangle^{obs}\right)}{\sigma_{\langle\tau\rangle}^2}\right]} \\
& \times \frac{1}{\sqrt{2\pi}\sigma_{\Sigma_*}}\exp{\left[-\frac{1}{2}\frac{\left(\Sigma_* - \Sigma_*^{obs}\right)}{\sigma_{\Sigma_*}^2}\right]},
\label{eqn:L}
\end{split}
\end{equation}

\noindent
where $\Theta=(r_s, \gamma, \rho_0)$, $\bar{\omega}_i$ is the measured angular velocity and $\sigma_{\bar{\omega},i}$ is the corresponding uncertainty, here $i$ runs over the radial RC bins.
The posterior from Bayes theorem is given by 

\begin{equation}\label{eqn:Bayes}
P(\Theta, \Sigma_*, \langle\tau\rangle|{\rm d,\mathcal{M}})=\frac{P(\rm{d}|\Theta, \Sigma_*, \langle\tau\rangle,\mathcal{M})P(\Theta, \Sigma_*, \langle\tau\rangle|\mathcal{M})}{P(\rm{d}|\mathcal{M})},
\end{equation}

\noindent
where $\mathcal{M}$ represents the assumed model and morphology (see Eq.~\ref{eqn:vel_model}), the likelihood
$P(\rm{d}|\Theta, \Sigma_*, \langle\tau\rangle,\mathcal{M})$ is given by Eq.~\ref{eqn:L}, the prior $P(\Theta, \Sigma_*, \langle\tau\rangle|\mathcal{M})$ is uniform in the variables and ranges indicated above, and the Bayesian evidence $P(\rm{d}|\mathcal{M})$ is an irrelevant normalization constant in this context. 

The posterior is sampled using the open source affine invariant Markov chain Monte Carlo (MCMC) ensemble sampler \texttt{emcee} \cite{2013PASP..125..306F}. In \texttt{emcee} we use 50 walkers and $2.5\times 10^5$ posterior samples (in the case of mock data we use $2\times 10^5$ posterior samples). We find it sufficient to discard the first 300 samples from each walker as burn-in. For all the runs, including different baryonic morphologies, the acceptance fraction varies between $0.28$ and $0.4$.

We further allow for the presence of residual undetected systematic effects between the different data sets (this is, between {\tt galkin$_{12}$} and the two Huang et. al.~\cite{2016MNRAS.463.2623H} data sets) that we are using in the observed RC compilation by using the procedure described in, e.g., \cite{Barnes+2003, Trotta+2011}. The procedure has the ability to detect systematic offsets in the data as well as underestimated measurement uncertainties. Specifically, it allows to check whether the reported error bars capture the intrinsic noise in the data.
For each data set $j$ and bin $i$, we rescale the variances of the binned data by a factor $1/\epsilon_j$, and replace the variance in the likelihood by  $\sigma^2_{ij}/\epsilon_j$. 
The nuisance parameters $\epsilon_j$ describe possible error rescaling due to undetected systematics in each data set. If there are no residual systematics, the observed variance in the RC is fully explained by the statistical error in our binned data and the data sets are mutually compatible (as they should be, for in compiling them we used an appropriate compatibility test), then $\epsilon_j \approx 1$. We expect residual systematics to yield $\epsilon_j \ll 1$, which would inflate the binned data variance to account for sources of variance not captured by the statistical errors.

The first term of the likelihood function in equation \eqref{eqn:L} needs to be modified to include the $\epsilon$ reascaling parameters. With this, the likelihood takes the form
\begin{equation}
    {\rm P}\left(\bar{\omega}^{ij}| \Theta, \Sigma_*, \langle\tau\rangle, \epsilon\right) = \frac{\sqrt{\epsilon_j}}{\sqrt{2\pi}\sigma_{\bar{\omega},ij}}\exp{\left[-\frac{1}{2}\frac{\left(\omega_c(r_i, \Theta, \Sigma_*, \langle\tau\rangle) - \bar{\omega}^{ij}\right)^2}{\sigma^2_{\bar{\omega}, ij}/\epsilon_j}\right]},
\label{eq:L_epsilon}
\end{equation}
where $i$ runs for the number of bins in data set $j$.

For the rescaling parameter $\epsilon_j$, following \cite{Trotta+2011} we use Jeffreys' priors which are uniform in $\log{\epsilon_j}$ and take the following form,
\begin{equation}
    {\rm P}(\log{\epsilon_j})=\left\{
                \begin{array}{ll}
                  2/3 \hspace{0.6cm} {\rm for -3/2 \leq \log\epsilon_j}\leq0\\
                  0   \hspace{1cm} {\rm otherwise}
                \end{array}
              \right.
\label{eq:prior_epsilon}    
\end{equation}
which correspond to a prior on $\epsilon_j$ of the form,
\begin{equation}
    {\rm P}(\epsilon_j)\propto1/\epsilon_j.
\end{equation}

We further investigate whether our analysis is robust with respect to the  $\sim$~20~kpc dip apparent in our data compilation but whose physical origin is unclear. In order to do so, we perform an alternative analysis excluding from the data sets bins between 10 and 20~kpc. The results are presented in Section~\ref{sec:results}.

\subsection{Performance of parameter reconstruction}
\label{sec:mocks}

\begin{table}
\centering

\begin{tabular}{ | c | c | c | c | c | c |}
\hline 

$r_s$ [kpc] & 5 & 8.5 & 12.5 & 16.5 & 20 \\
\hline
$\gamma$ & 0.1 & 0.4 & 0.8 & 1.2 & 1.5 \\ 
\hline
$\rho_0$, $\rm{[GeV/cm^3]}$ & \multicolumn{1}{c}{} & \multicolumn{3}{c}{0.4} & \\
\hline 
$\Sigma_*$, $\rm[M_{\odot}/kpc^2]$ & \multicolumn{1}{c}{} &\multicolumn{3}{c}{$38\times10^6$}& \\
\hline 
$\left<\tau\right>$ & \multicolumn{1}{c}{} &\multicolumn{3}{c}{$2.17\times10^{-6}$}& \\
\hline
\end{tabular}
\caption{\label{tab:par space}Fiducial parameter values used in mock RC data generation. We use every possible combination of the parameters in the table, producing 25 fiducial points in our 5 dimensional parameter space (see Section~\ref{sec:mocks} for details).}
\end{table}

\begin{figure}
\begin{centering}
\resizebox{10cm}{9cm}
{\includegraphics{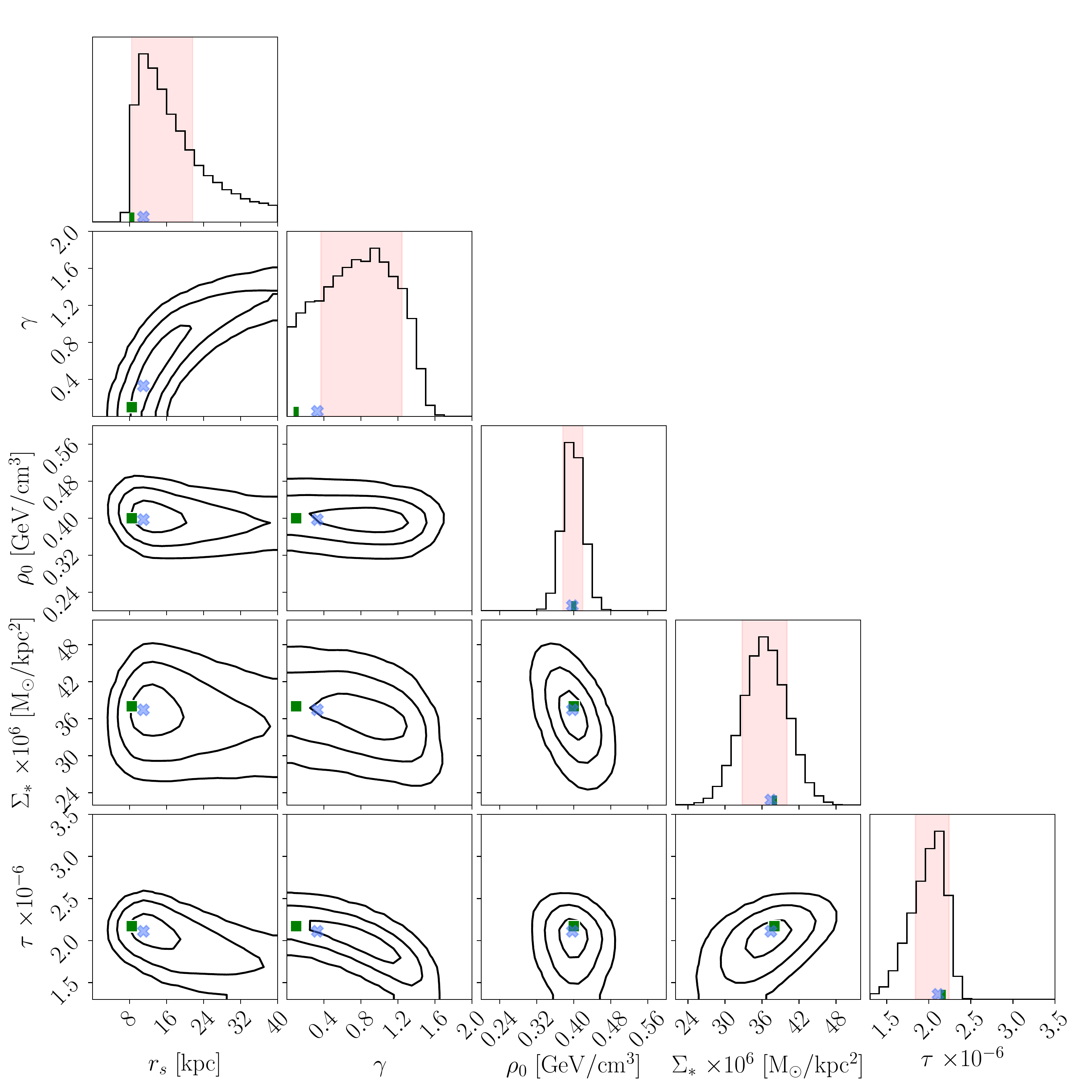}}
\resizebox{10cm}{9cm}
{\includegraphics{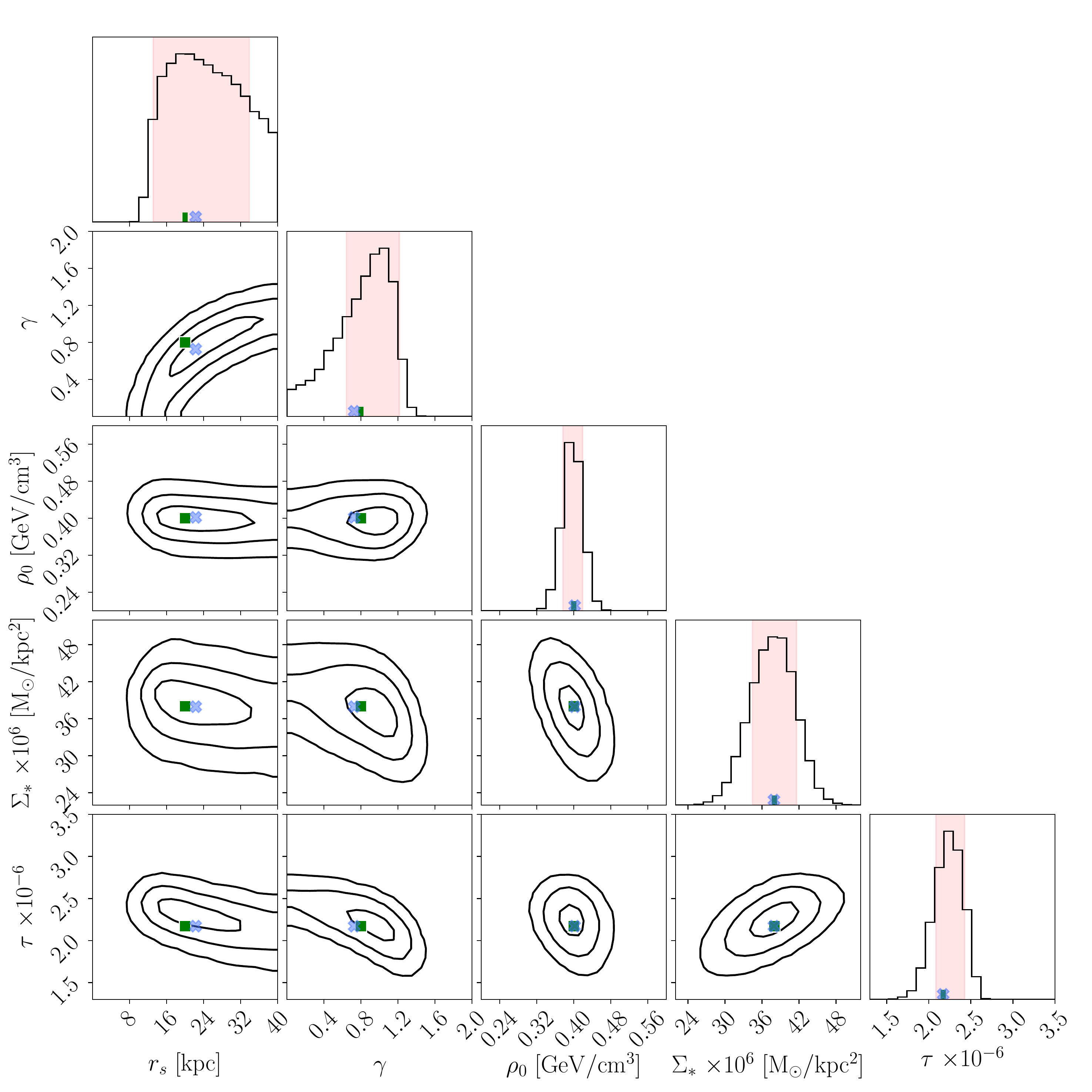}}
\caption{\label{fig:corner}{\it Parameter reconstructions for mock data}. Two example reconstructions of the 5 parameters in our model obtained from simulated data sets. Posterior marginalized 2D and 1D distributions are shown for 2 out of 25 fiducial points in parameter space. Green squares indicate the true fiducial values used to generate the data, while blue crosses show the maximum likelihood values. The contours delimit regions of 68\%, 95\%, 99\% probability, while red shaded areas show the 68\% highest posterior density (HPD) credible intervals.}
\end{centering}
\end{figure}

 We wish to investigate the sampling errors of our parameter reconstruction procedure (i.e. over repeated data realizations) in order to establish, for example, coverage properties of our Bayesian posterior regions under different procedures for creating them. As these properties in general depend on the true value of the parameters, we investigate 25 different choices for such fiducial parameter values, as given by the possible combinations in Table~\ref{tab:par space}. For each set of fiducial parameters, we generate 100 mock data sets, drawn from the likelihood given by Eq.~\eqref{eqn:L}, with measurement error in each bin equal to the estimated standard deviation within that bin. The underlying angular velocity are obtained using Eq.~\ref{eqn:vel_model}, where  $r_s,\gamma,\rho_0, \Sigma_*$, and $\left<\tau\right>$ are set to the {\it fiducial} values. For the baryonic components we adopt the HG disk~\cite{Han:2003ws}, E2 bulge~\cite{Dwek:1995xu}, and gas~\cite{Ferriere:2007yq,0004-637X-497-2-759} as our fiducial reference morphologies (see discussion in Section~\ref{sec:baryons}).

\section{Results}  
\label{sec:results}

\subsection{Mock data reconstruction}
\label{sec:mock data}
Using the priors described in Section \ref{sec:MCMC}, we have fit 100 mock RCs for each fiducial point in the parameter space in order to test our reconstruction procedure. Figure~\ref{fig:corner} shows, as a first example, the full posterior distribution given mock data generated by two different choices of the fiducial parameters. In both cases, the local dark matter density $\rho_0$ is very well recovered while the inner slope of the density profile $\gamma$ and the scale radius $r_s$ are highly degenerate. These features are present in the posteriors for all 25  fiducial points.

 We further estimate the accuracy of the reconstruction for the dark matter parameters of interest: $r_s$, $\rho_0$, and $\gamma$. In order to do so we define the `fractional standard error' where in the definitions below $n_i$ is the number of mock data realizations for each fiducial point in the parameter space, $\theta_\mathrm{true}$ is the value of the fiducial parameters used and $\hat{\theta}$ is a point estimate. We consider four such point estimates: the mean, median, maximum a posteriori (MAP), and the maximum likelihood (ML). These estimates (except for the ML, which is obtained directly from the likelihood) are derived from the marginal 1D posterior distribution for the parameter in question.

The fractional standard error (FSE) is defined as the square root of the mean squared error normalized by the true value: 
$$\mathrm{FSE}=\frac{\sqrt{\frac{1}{n_i}\sum_{i=1}^{n_i}(\hat{\theta}_i-\theta_\text{true})^2}}{\theta_\text{true}}.$$
The FSE is a measure of the (fractional) precision and accuracy of the reconstructed point estimate. 

The detailed plot for the FSE with respect to different estimators for each of the 3 parameters is shown on Fig.~\ref{fig:FSE} of Appendix~\ref{App:mock data} where it is presented in form of color maps.
We can summarize these results as follows:

\begin{itemize}
\item The values for the FSE of $\rho_0$ varies between 0.03 and 0.06 (see Fig.~\ref{fig:FSE}) 
 independently of the chosen point estimator. This indicates that the recovered value of the local dark matter density can be expected to have better than $6\%$ precision for any reasonable values of the dark matter halo parameters.
\item The accuracy and precision of the $\rho_0$ estimators slightly improve for larger values of $\gamma$ (except for the ML).
\item The accuracy and precision of the estimators for $\gamma$ and $r_s$ (except for MAP and ML estimators of $r_s$) is better for higher values of $\gamma$ and $r_s$.
\item In general the estimates of $\gamma$ and $r_s$ have poor accuracy. This is due to the curved degeneracy between these two parameters as illustrated in Fig.~\ref{fig:corner}. While a particular combination of $\gamma$ and $r_s$ may be constrained, marginalizing the posterior to 1D yields poor estimates of these quantities individually.
\end{itemize}



We now turn our attention to the coverage properties of the reconstructed intervals for each parameter. Coverage is the fraction of the time, over repeated data realizations, that the inferred interval contains the true value of the parameter. We consider different definitions for our 1D intervals: the highest posterior density (HPD) interval from the 1D marginal posterior (which by definition is the shortest credible interval containing a given fraction of the total probability); median credible regions (or equal-tailed interval) (CR), that are defined as the 16th and 84th percentile of the posterior; and the profile likelihood\footnote{The profile likelihood function is defined for each value of $\Theta$ by maximization of the likelihood function over the nuisance parameters.} interval (PL), the region where the log-likelihood is within 1 of its maximum value (this is a classical way of constructing confidence intervals, based on Wilks' theorem~\cite{wilks1938}).

Coverage plots from 100 mock data realization as a function of the assumed fiducial parameter values are presented in Fig.~\ref{fig:HPD} of Appendix~\ref{App:mock data}. These results can be summarized as follows:

\begin{itemize}
    \item In the case of the local density $\rho_0$, all intervals overcover (i.e., they are conservative), with the most severe overcoverage occurring for the PL interval (bottom left panel of Fig.~\ref{fig:HPD}).
    \item For the scale radius $r_s$ the situation is similar to the previous point, but the HPD and CR Bayesian intervals suffer from undercoverage in some parts of the parameter space, notably for small values of $\gamma$ and more severely so for the CR intervals.
    \item For $\gamma$ we observe both over- and under-coverage for the Bayesian intervals (both HPD and CR), while the PL interval always overcover.  
\end{itemize}
\noindent

\begin{figure}
\includegraphics[width=0.9\textwidth]{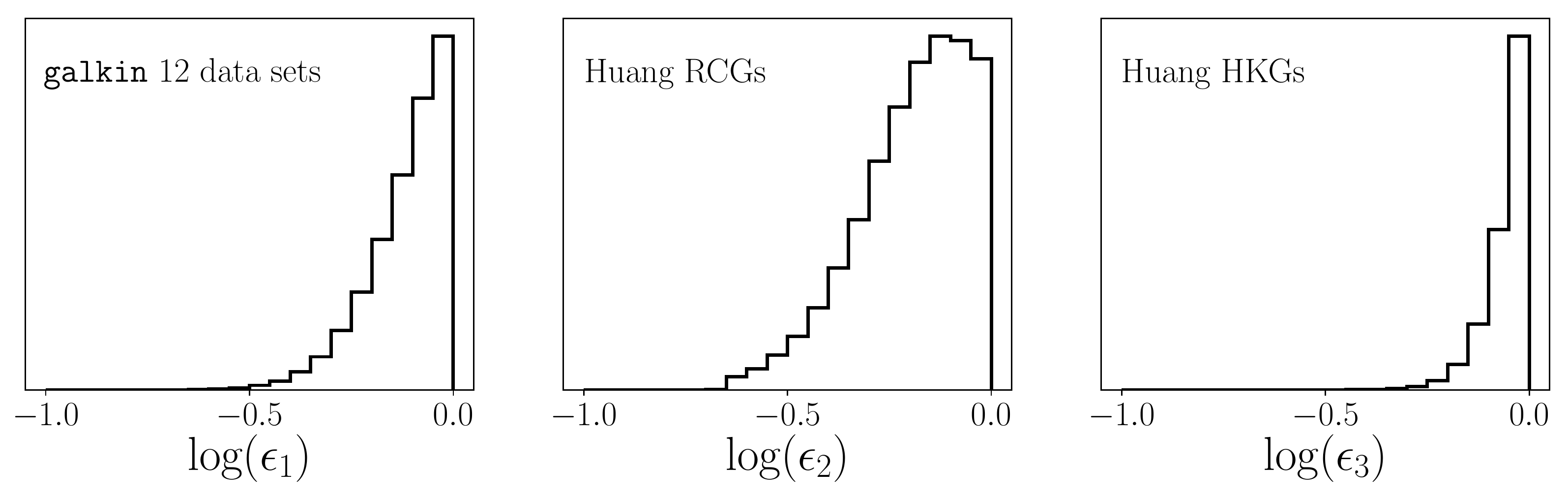}
\centering
\caption{ {\it Based on real data.} One-dimensional marginalized posterior probability distributions for the error bar rescaling parameters. The peaks around $\log(\epsilon_j)=0$ indicate that no error bar rescaling is necessary, which would otherwise have been the case had systematic errors been present among the data sets considered. \label{fig:case3_epsilon}}
\end{figure}

The reasons for the behaviours described in the last two points can be traced back to the 2D degeneracy between $r_s$ and $\gamma$: once marginalized to 1D the resulting 1D Bayesian intervals can have poor coverage depending on where in the 2D degenerate region the true value is located. The over-coverage observed for all the fiducial parameter values for the PL is a consequence of the PL intervals being typically larger than both the Bayesian ones. For this reason, for $\gamma$ and $r_s$ the 1D coverage of Bayesian intervals can stray from being exact, while being much better behaved in the 2D plane spanning both. We leave further investigation of this effect for future work. 
Notice that the two nuisance parameters are always well recovered within the observed $1\sigma$ uncertainties, independently of where we are in the parameter space, as they are constrained by an independent piece of the likelihood.

\begin{figure}
\centering
\includegraphics[width=0.6\textwidth]{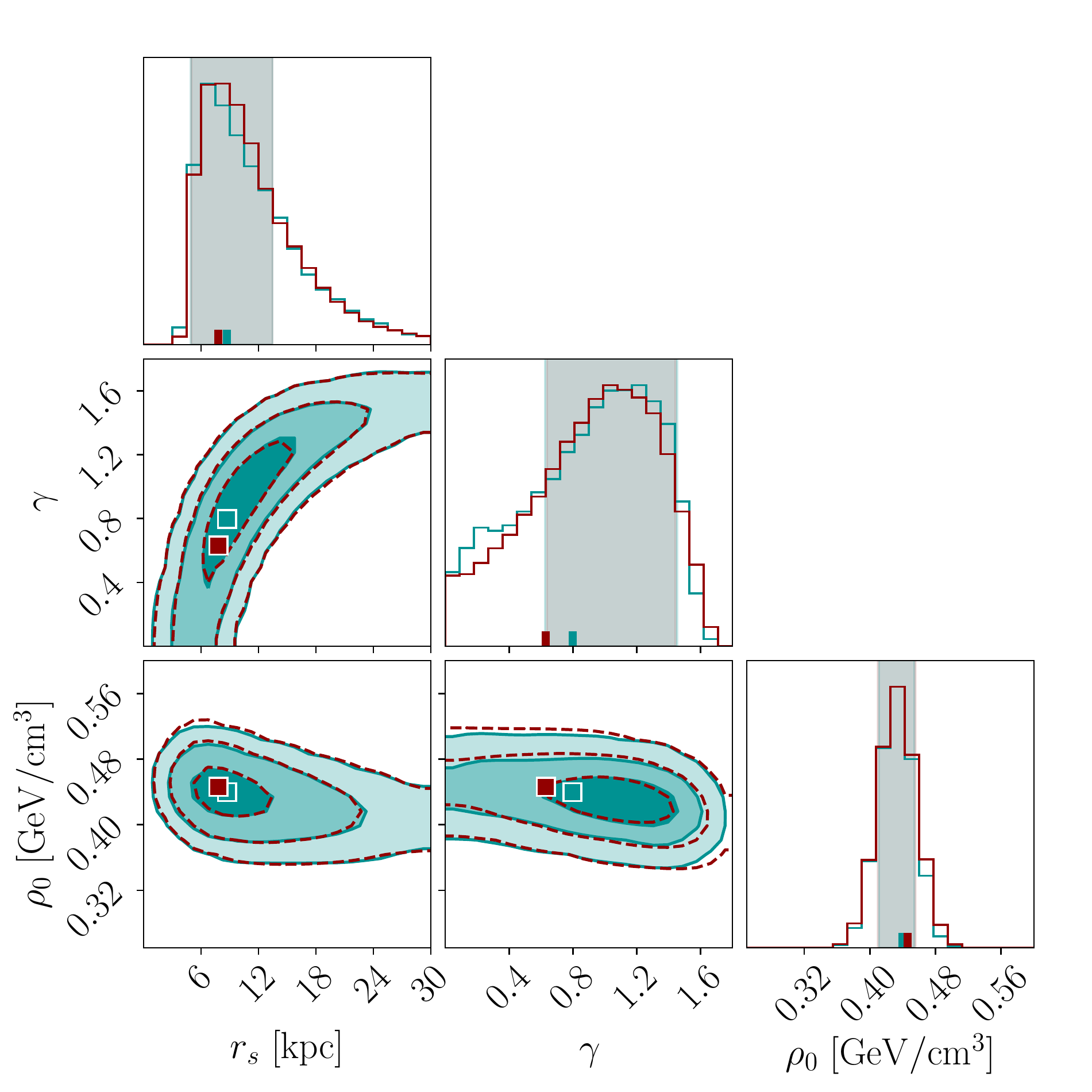}
\caption{\label{fig:epsilon_corner} {\it Based on real data}. One and two-dimensional marginalized posterior distributions for dark matter parameters. In teal (dark red) are the results without (with) the inclusion of  error bar rescaling parameters for \texttt{galkin}$_{12}$ and Huang~et~al.~\cite{2016MNRAS.463.2623H} data. The contours delimit regions of 68\%, 95\%, 99\% probability, while light teal and light red bands in the 1D plots indicate 68\% HPD region for each analysis. The resulting inference is robust, with the analysis including error rescaling parameters giving only slightly wider regions, compatible with the case where all $\epsilon_j$ are fixed to 1. The squares in 2D plots and dashes in 1D plots indicate the maximum likelihood values in each case.}
\end{figure}

\subsection{Reconstruction of parameters using real data \label{sec:real data}}
In this section we present the inference results when using the data sets for the Milky Way presented in Section~\ref{sec:observed RC}. We have argued above that our Bayesian procedure for selecting compatible data sets robustly eliminates data sets that are likely to be systematically offset. An inspection of the \texttt{galkin}$_{12}$ compilation selected in this way shows that the data within each bin are nicely Gaussian distributed around their mean value. This is another indication of compatibility, and it validates our usage of a Gaussian likelihood function. Here, we further investigate the possibility of remaining undetected systematic effects within our compilation of RC data by applying the Bayesian formalism explained in Section~\ref{sec:method}, where we introduced the variance scaling parameters $\epsilon_j$ (see Eq.~\ref{eq:L_epsilon}).

\begin{figure}
\centering
\includegraphics[width=0.7\textwidth]{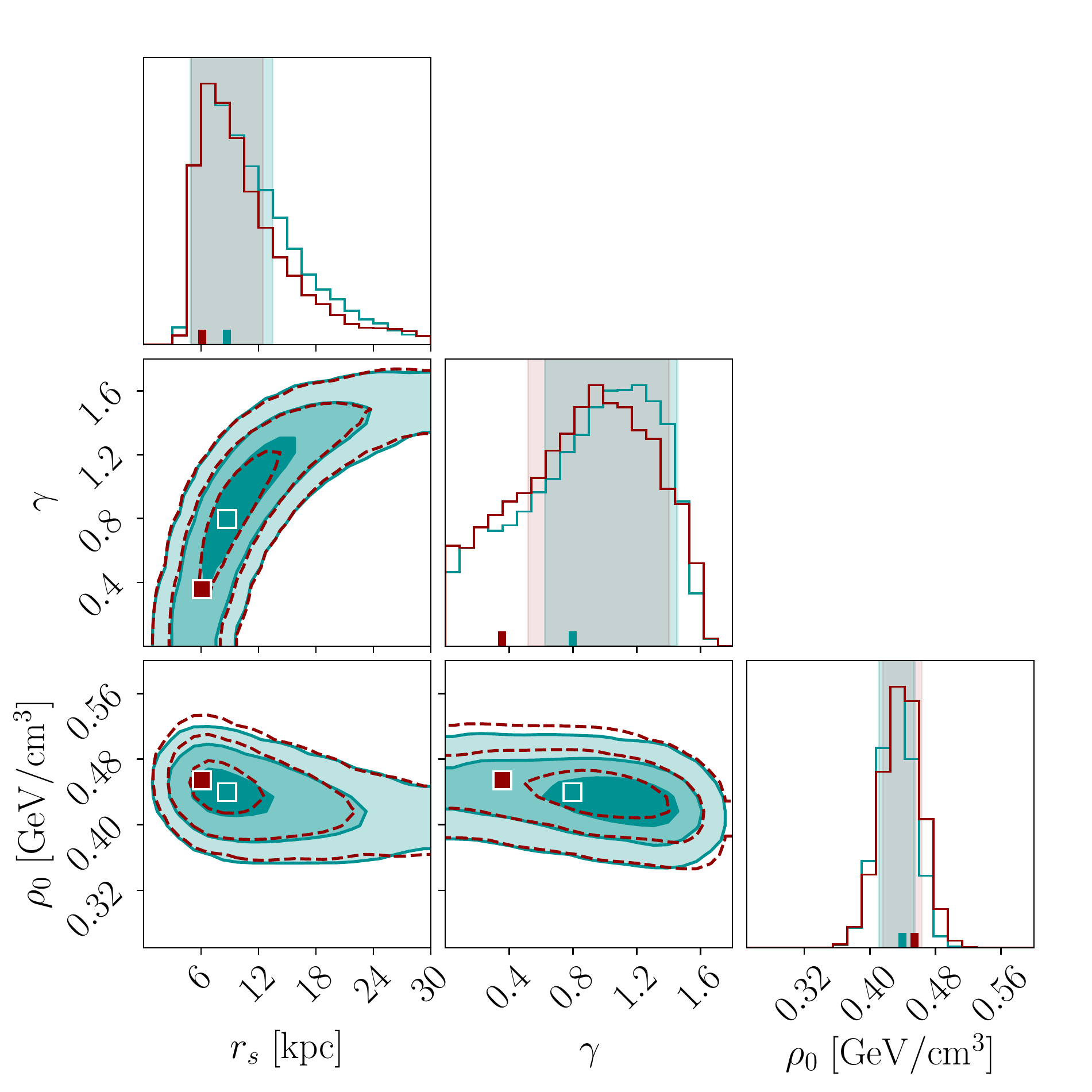}
\caption{{\it Based on real data}. One and two-dimensional marginalized posteriors for dark matter parameters. In teal are the results of the standard analysis. In dark red, we show the results when bins between 10 and 20~kpc are omitted from the likelihood. The reconstruction is therefore largely insensitive to dip in the rotation curve between 10 and 20~kpc. Light teal and light red bands indicate 68\% HPD regions for the standard analysis and the analysis when removing bins between 10 and 20 kpc, respectively. The squares in 2D plots and dashes in 1D plots indicate the maximum likelihood values in each case.
\label{fig:missing_points}}
\end{figure}


\begin{figure}
\centering
\includegraphics[width=0.8\textwidth]{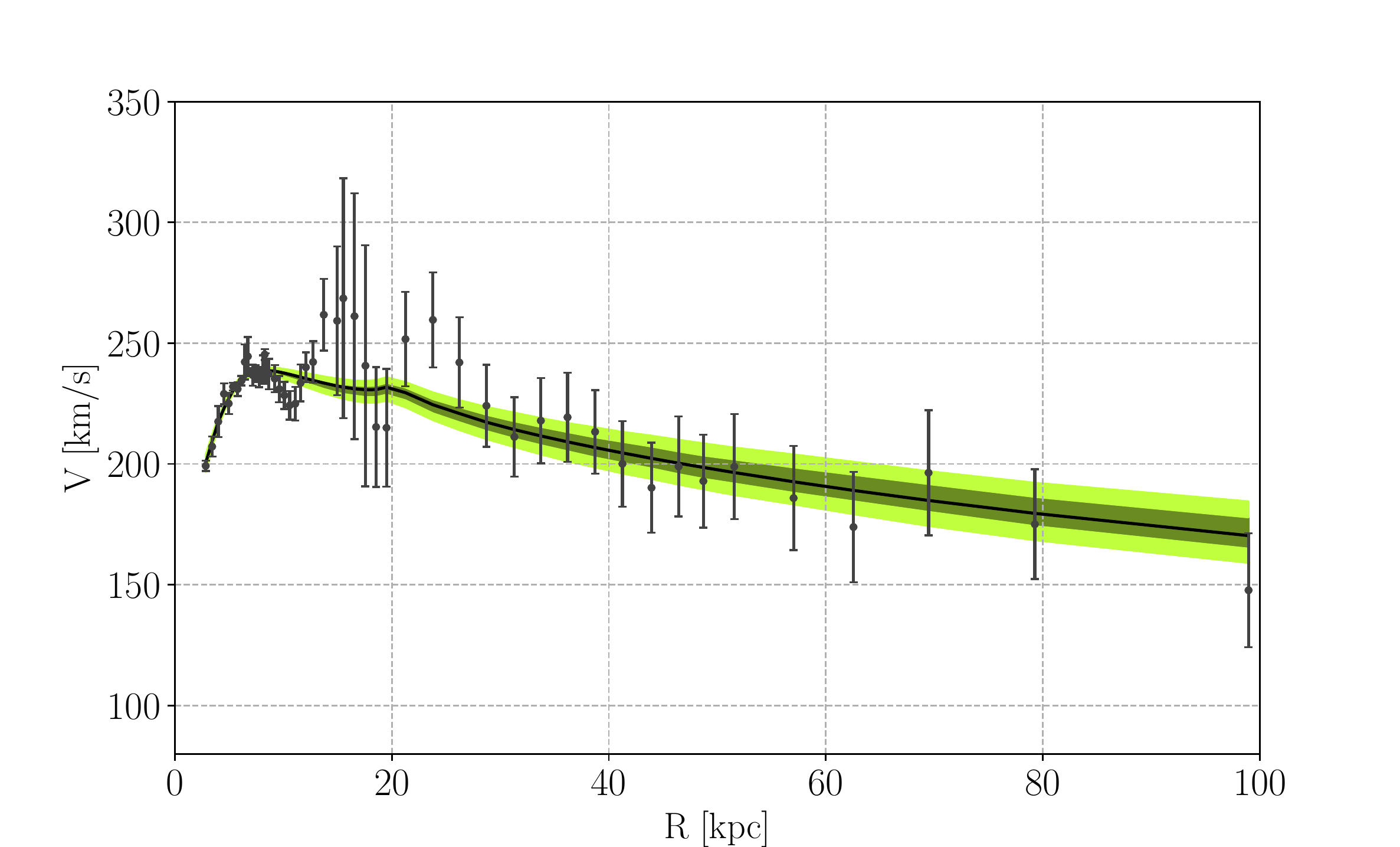}
\caption{\label{fig:RC_HPD} {\it Based on real data.} Black solid line indicates the RC using the maximum likelihood parameter values for our reference baryonic morphology. Dark green and light green bands show the RC corresponding to 68\% and 95\% credible region (HPD) in our parameter space (dark matter + baryonic nuisance parameters), using our reference baryonic morphology.}
\end{figure}

We look for systematic differences between the \texttt{galkin}$_{12}$ and the two data sets of Huang~et~al.~\cite{2016MNRAS.463.2623H} by introducing three epsilon scaling parameters, one for each data set. Notice that if there were any systematic errors within the \texttt{galkin}$_{12}$ data, the corresponding $\epsilon_j$ parameter would prefer values smaller than unity to compensate for them. The 1D marginal posterior distributions for the $\epsilon_j$ parameters are shown in Fig.~\ref{fig:case3_epsilon}, where we see that the posteriors peak at $\log(\epsilon_j) = 0$ for all three parameters, thus indicating that there is no need to further increase their uncertainties. In the following, we thus remove the $\epsilon_j$ parameters from the analysis (having checked that the ensuing posterior distributions for the parameters of interest do not exhibit any significant shift when including or excluding the $\epsilon_j$ parameters). In Fig.~\ref{fig:epsilon_corner} we also show the comparison of the MCMC analysis with and without including error rescaling parameters. This plot indicates practically no difference in the resulting posteriors.

We now check whether the $\sim 20$~kpc dip in the data (see Fig.~\ref{fig:RCs1}) affects the parameter reconstruction. In order to do so, we re-run the MCMC analysis (without including error rescaling parameters) for both the whole binned RC data and omitting the bins between 10 and 20~kpc. The results of these two runs are compared on Fig.~\ref{fig:missing_points}, where we see essentially no difference in the resulting posteriors (the ML point does however suffer a shift, negligible in the case of $\rho_0$). This exercise, together with the fact mentioned in Section~\ref{sec:bayesian_evidence} that binned {\it whole} \texttt{galkin} data compilation once combined with the outer RC data show almost no change in the resulting Bayesian inference with that of \texttt{galkin}$_{12}$ + extended Huang~et~al. RC, indicates the importance of the RC data beyond the $\sim$~20~kpc. 


Finally in Fig.~\ref{fig:RC_HPD} we superimpose RC based on the maximum likelihood parameter values over the observed RC data, and indicate the 68\% and 95\% HPD regions from our analysis. The latter is derived from the corresponding pdf of the circular velocity which, in turn, is calculated based on the pdfs of the \texttt{emcee} samples.

\begin{figure}
\centering
\includegraphics[width=0.8\textwidth]{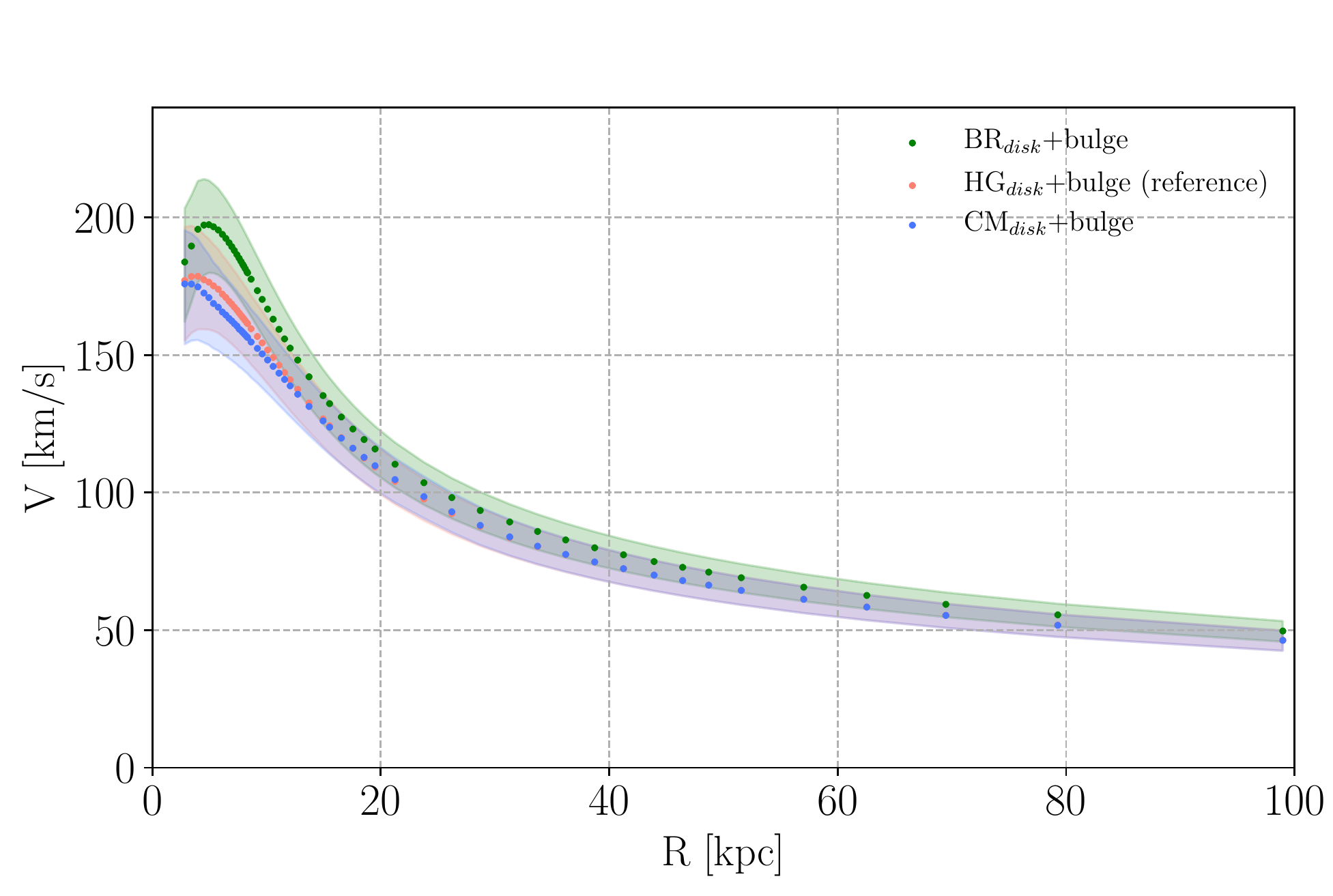}
\caption{\label{fig:disk_morphologies}{\it Based on real data.} Sum of the circular velocities of the disk and bulge, adopting different disk morphologies (see text for more details). The measured values of the surface stellar density and the microlensing optical depth have been included in the likelihood, taking into account the quoted uncertainties: $\Sigma_*^{obs}=(38\pm 4)\times10^6\hspace{0.1cm}\rm{M_{\odot}/kpc^2}$ and to $\langle\tau\rangle^{obs}=(2.17\pm{0.47})\times 10^{-6}$.}	 
\end{figure}

The quality of the best-fit RC when adopting different morphologies is presented in Appendix~\ref{App:morphologies}, in terms of its reduced chi-squared (Table~\ref{tab:chi2_ML_allmorph}), showing a good quality of fit for all morphologies. Instead, here we summarize details of how the Bayesian posterior inferences change when adopting three different and critical disk morphologies. In particular, we make use of two additional disk morphologies  (demonstrated in Fig.~\ref{fig:disk_morphologies})  that are obtained with the lightest disk (also referred to as CM~\cite{2011MNRAS.416.1292C}) and heaviest disk (also referred to as BR~\cite{2013ApJ...779..115B}) so as to present a ``bracketing'' of the actual systematic uncertainty due to ignorance of the actual morphology of the visible component. Notice that we keep the bulge morphology fixed and the same as in our reference morphology\footnote{If we vary bulge morpholgies while keeping a disk morphology fixed, the resulting dark matter parameters remain almost unchanged with mild variations well within the corresponding 68\% HPD regions, see Appendix~\ref{App:morphologies}.}.

\begin{figure}
\centering
\includegraphics[width=0.7\textwidth]{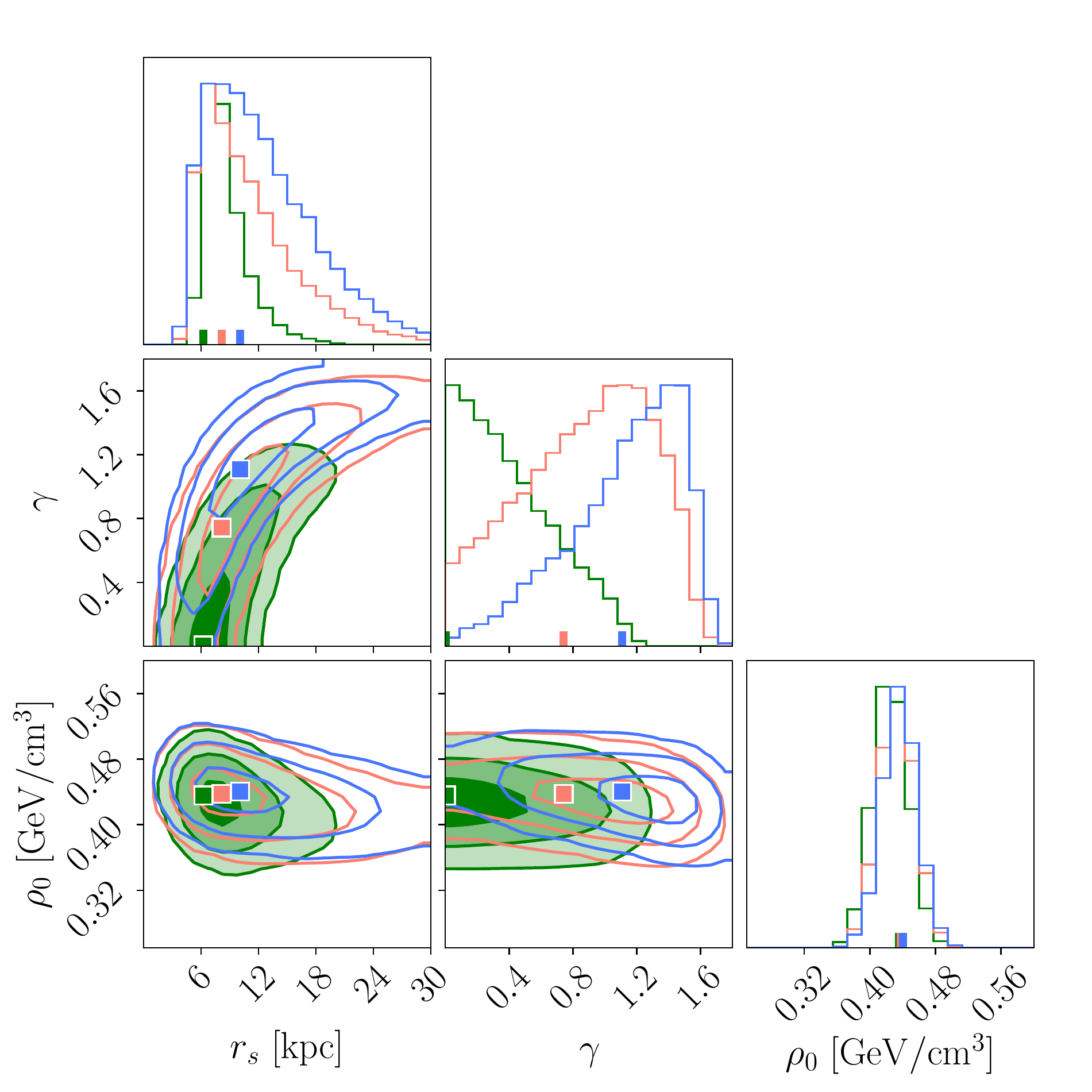}
\caption{\label{fig:three morphologies}	 {\it Based on real data.} One and two-dimensional marginalized posterior distributions for the dark matter parameters, comparing the results for different assumptions about baryonic morphology: the heaviest disk BR~(\cite{2013ApJ...779..115B}, green), the average disk HG (our reference disk~\cite{Han:2003ws}, pink) and the lightest disk CM~(\cite{2011MNRAS.416.1292C}, blue). The squares in 2D plots and dashes in 1D plots indicate the maximum likelihood values in each case.}
\end{figure}

\begin{table}
\begin{tabular}{ | c | c | c | c | c | c | c | c |}
\hline
 Disk morph.    & ML $r_s$  & ML $\gamma$ & ML $\rho_0$ & HPD$_{68}$ $r_s$  & HPD$_{68}$ $\gamma$ & HPD$_{68}$ $\rho_0$ & $\chi^2_\mathrm{red}$ \\
\hline 
\hline
$\rm{BR}$~\cite{2013ApJ...779..115B} &6.4 & 0.01  & 0.43 & $7.1^{+2.3}_{-1.1}$ & <0.93 (95\%) & $0.42^{+0.02}_{-0.02}$ & 0.91 \\
\hline
HG~\cite{Han:2003ws}& 8.7 & 0.80 & 0.44 & $6.3^{+7.2}_{-1.4}$ & $1.20^{+0.25}_{-0.58}$ &$0.44^{+0.02}_{-0.03}$ & 0.94 \\ 
\hline
CM~\cite{2011MNRAS.416.1292C}  & 10.1 & 1.11 & 0.44 &
$7.9^{+6.9}_{-3.0}$ & $1.44^{+0.16}_{-0.48}$ &$0.44^{+0.02}_{-0.02}$& $1.01$\\ 
\hline

\end{tabular}
\caption{\label{tab:chi2_ML} The maximum likelihood (ML) and the maximum a posteriori (MAP) estimates with uncertainties obtained from the 68\% HPD region (except for $\gamma$ in the BR morphology case, where we indicate the 95\% upper limit as the 1D marginal posterior is compatible with~0). We also give the reduced $\chi^2$ values for the best fit (i.e. ML) parameters in the last column.}
\end{table}

The results of the MCMC analysis for the three morphologies are presented on Fig.~\ref{fig:three morphologies} and summarized in Table~\ref{tab:chi2_ML}. In this table we quote the reduced $\chi^2$ that is calculated for 49 degrees of freedom (54 bins - 5 free parameters).
From this figure and the table it is evident, that whilst posteriors on $\gamma$ and $r_s$ do change with different morphologies, the local density is unchanged with a maximum likelihood value of $\sim 0.44\,\rm{GeV/cm^3}$. We also find that it is impossible, at the moment, to rule out any of them using the goodness of the fit. Finally, we emphasize that in our analysis we fix the Sun's galactocentric distance, whose exact value might impact the determination of the dark matter distribution.  For example, reference \cite{Benito:2019ngh} shows that there is a correlation between the local dark matter density and the value of the Sun's galactocentric distance. However, \cite{Benito:2019ngh} only makes use of the inner Milky Way RC data (differently from us), and adopts a different statistical approach than ours. Furthermore, \cite{Benito:2019ngh} does not assess the compatibility of the \texttt{galkin} sub-sets of data, as we do here, and use instead the original \texttt{galkin} compilation. Therefore, the impact of varying the Sun's galactocentric distance on the determination of the local dark matter density in our study cannot be directly compared with what obtained in \cite{Benito:2019ngh}. Despite the differences in the methodology, we notice that our value of the local dark matter density for the Sun's galactocentric distance of 8.34~kpc is compatible within 95~\% HPD region with the result of~\cite{Benito:2019ngh}. 

Although our derived value of the local dark matter density is in good agreement with recent measurements (see e.g. \cite{2017MNRAS.465.1621P,Sivertsson:2017rkp} and references therein)\footnote{Notice, that when we apply our analysis with our reference baryonic morphology to the recently published Gaia DR2 data \cite{2019ApJ...871..120E}, the local dark matter estimate is compatible with our estimate (when using \texttt{galkin$_{12}$} data assuming the local standard of rest and the values of $\rm V_0,\,R_0$ as in \cite{2019ApJ...871..120E}) and with the estimate from \cite{deSalas:2019pee} when the authors use their B2 baryonic model. Ref. \cite{deSalas:2019pee} appeared in arXiv while our paper was submitted to this Journal.}, it is in tension with the value derived in \cite{2016MNRAS.463.2623H}. The reason for that is the difference in the modelling of the total Milky Way  gravitational potential between our work and \cite{2016MNRAS.463.2623H}. In particular, in the latter the authors add two additional components (with respect to ours) in order to model a hypothetical caustic ring of dark matter in the Galactic plane. Doing so the derived value of the local dark matter density (not accounting for that in the ring) decreases by $\sim~25$~\%. In order to derive this number we have included the dark matter ring in our model in the same manner as in \cite{2016MNRAS.463.2623H}, we then fixed all the related parameters to their best-fit values and we obtained the value of the local dark matter density consistent within 68~\% HPD region with that of \cite{2016MNRAS.463.2623H}.

\section{Conclusions \label{sec:conclusions}}
We have developed a novel Bayesian methodology to reliably and precisely infer the distribution of dark matter within the Milky Way using the rotation curves method. Based on the most recent data sets for the rotation curve, we first selected a subset of tracers that are mutually compatible with each other, thus excluding the ones that {\it may} be suffering from systematic bias. We then demonstrated the statistical performance of our rotation curve method (over many data realizations) by applying it to simulated data, which we generated respecting the statistical properties of the observed rotation curve. We generalized our procedure against (systematic) uncertainties on the visible component of the Milky Way, by applying it to different possible baryonic morphologies, in order to bracket the current uncertainties on the shape and normalization of the Galactic stellar disk(s) and bulge, as well as of the (subleading) interstellar gas component (see Section~\ref{sec:real data} and Appendix~\ref{App:morphologies}).

Our most relevant findings can be summarized as follows:

\begin{itemize}
\item Based on the kinematic data and the Galactic parameters used in this work, we infer a local dark matter density is $\rho_0=0.43\pm{0.02}\pm{0.01}\,\rm{GeV/cm^3}$, where the first errors are statistical and the second systematic (adopting different baryonic morphologies). This number is in fair agreement with previous determinations in the literature (see e.g. \cite{Catena:2009mf,Iocco:2011jz,2015JCAP...12..001P}).
\item The determination of $\rho_0$ is very precise (68~\% credible region always between 4\% and 7\%), and extremely accurate: when tested against mock data, the value of $\rho_0$ inferred through the procedure does not fall more than $\sim$6\% away from the true value. This is of course predicated on our model for the RC data to be a complete and accurate representation of how the true data are generated, in that the reconstruction accuracy is only valid for mock data generated from the model itself.
\item In the gNFW formulation, the scale radius $r_s$ and the slope of the inner density profile $\gamma$ are degenerate, thus making the separate reconstruction of the two parameters more challenging. Whereas the combination of the two is relatively well constrained in a 2-dimensional space, the individual posteriors, once marginalized to 1D, are not very informative. 

\item The last two statements are robust against uncertainties in the baryonic morphologies (i.e. they are unaffected by the systematic ignorance of the baryonic morphology), and the shape of the underlying dark matter profile (precision and accuracy are little sensitive of the parameter space region where the underlying true value lies).
\end{itemize}

From the above, we conclude that the inference of the local dark matter density $\rho_0$ is extremely robust: it is important to stress that the central value obtained with our method is both very precise (with the 68\% credible interval being $\sim$~4\%) and 
highly accurate (never falling farther than $\sim$~6\% from the true value in simulated data). We also found that the inferred value and error bars are stable with respect to the morphology of the visible, baryonic component (with $< 3\%$ shift in the inferred $\rho_0$). Our results are also stable with respect to various tests for the presence of systematic errors in the RC data used. 

The methodology presented here is very flexible, as it allows the effortless incorporation of further data sets and accompanying nuisance parameters. We therefore expect that it will prove very useful in analyzing current and upcoming data about the structure and properties of the Milky Way mass distribution. 

\section*{Acknowledgements}
 We thank the referee for his/her comments that helped improve the paper.
 EK work was supported by the S\~ao Paulo Research Foundation (FAPESP) under Grant No. 2016/26288-9 and by the grant AstroCeNT: Particle Astrophysics Science and Technology Centre is carried out within the International Research Agendas programme of the Foundation for Polish Science co-financed by the European Union under the European Regional Development Fund. F.~I. acknowledges support from the Simons Foundation and FAPESP process 2014/11070-2. AG-S and RT are supported by Grant ST/N000838/1 from the Science and Technology Facilities Council (UK). RT was partially supported by a Marie-Sklodowska-Curie RISE (H2020-MSCA-RISE-2015-691164) Grant provided by the European Commission.  
This work has been possible through the FAPESP/Imperial College London exchange grant ``SPRINT'', process number 2016/50006-3. This research was supported by resources supplied by the Center for Scientific Computing (NCC/GridUNESP) of the S\~ao Paulo State University (UNESP).

\appendix
\renewcommand\thefigure{\thesection.\arabic{figure}}  

\section{Mock data reconstruction: accuracy and precision}
\label{App:mock data}
\begin{figure}
\resizebox{16.1cm}{4.4cm}
{\includegraphics{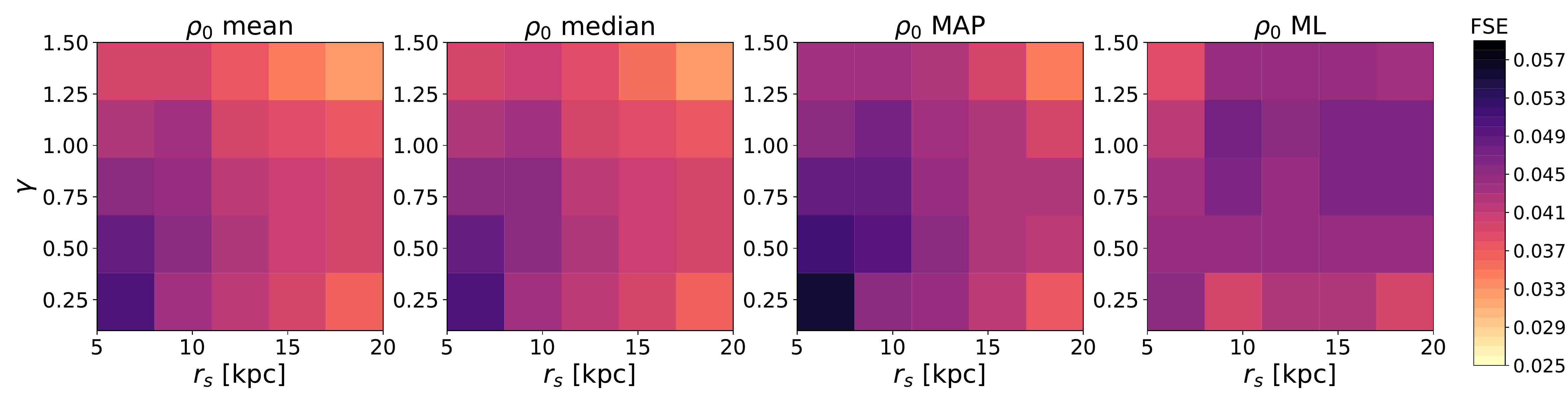}}
\resizebox{16.2cm}{4.2cm}
{\includegraphics{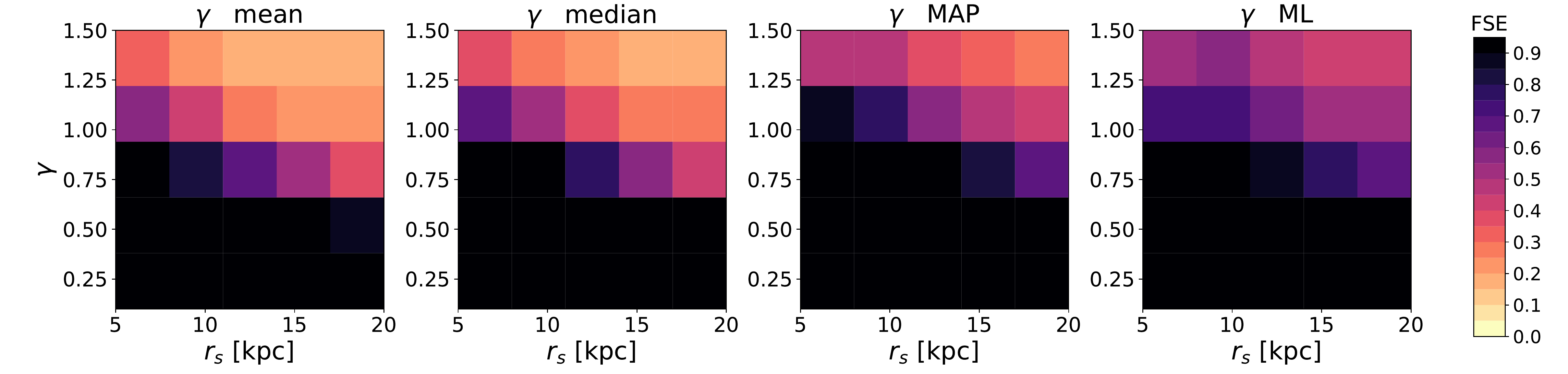}}
\resizebox{16.2cm}{4.4cm}
{\includegraphics{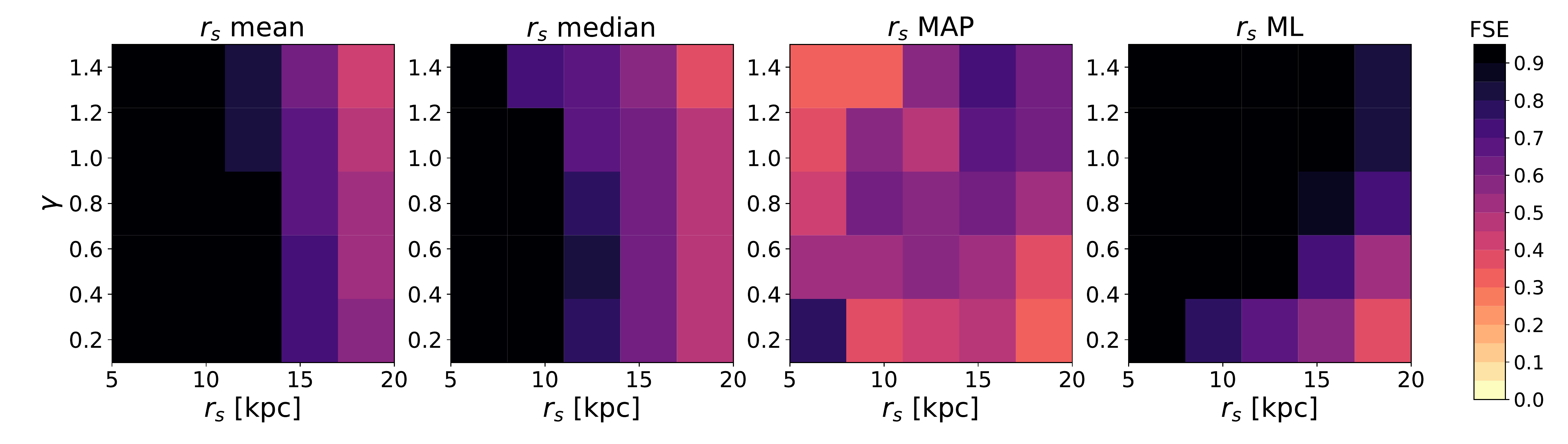}}
\caption{Fractional standard error (FSE) of reconstruction of local dark matter density $\rho_0$, inner slope of the density profile $\gamma$, and the scale radius of the dark matter halo $r_s$ (top to bottom) when using different point estimates (from left to right: mean, median, maximum a posteriori, and maximum likelihood). Note the difference in the color bar scale for the plots relative to that for estimators of $\rho_0$ (first row).\label{fig:FSE}}
\end{figure}

On Fig.~\ref{fig:FSE} we demonstrate the fractional standard error from 100 mock data realization as a function of the assumed fiducial parameter values, where the four columns correspond to the four different definitions of point estimate, and rows correspond to three parameters of interest. The color bars next to the plots decode the corresponding values of FSE. The results of this figure are summarized in Section~\ref{sec:mock data}.

\begin{figure}
 \begin{centering}
\resizebox{13.4cm}{4.7cm}
{\includegraphics{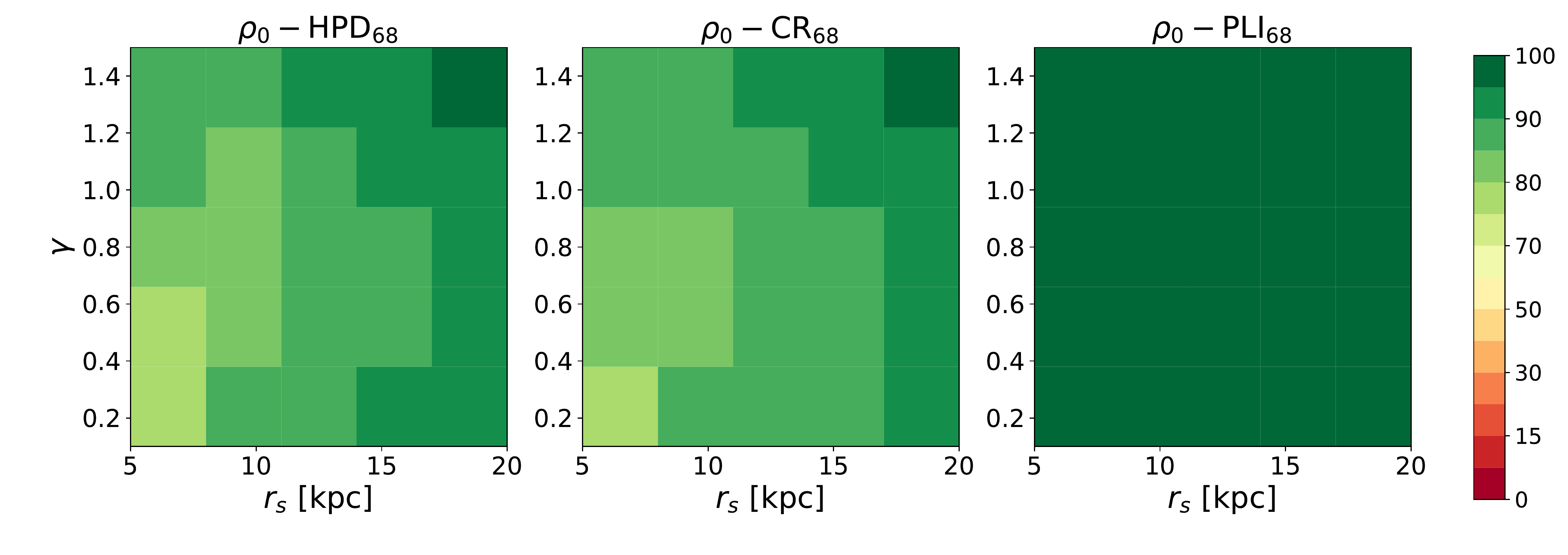}.pdf}
\resizebox{13.2cm}{4.5cm}
{\includegraphics{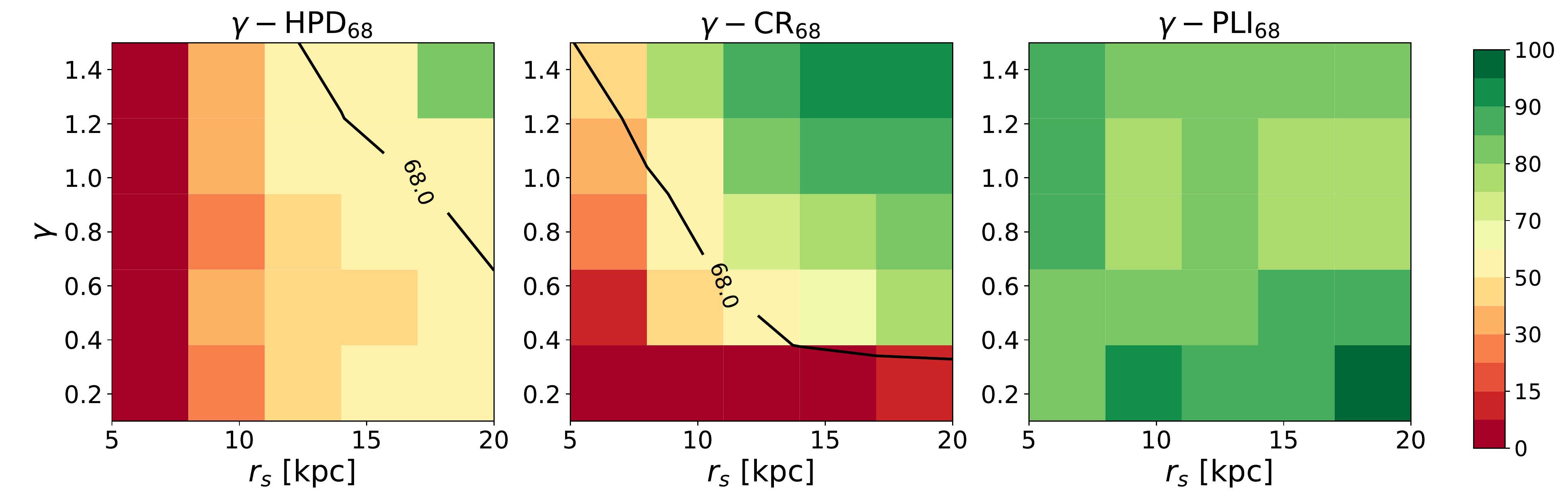}.pdf}
\resizebox{13.2cm}{4.5cm}
{\includegraphics{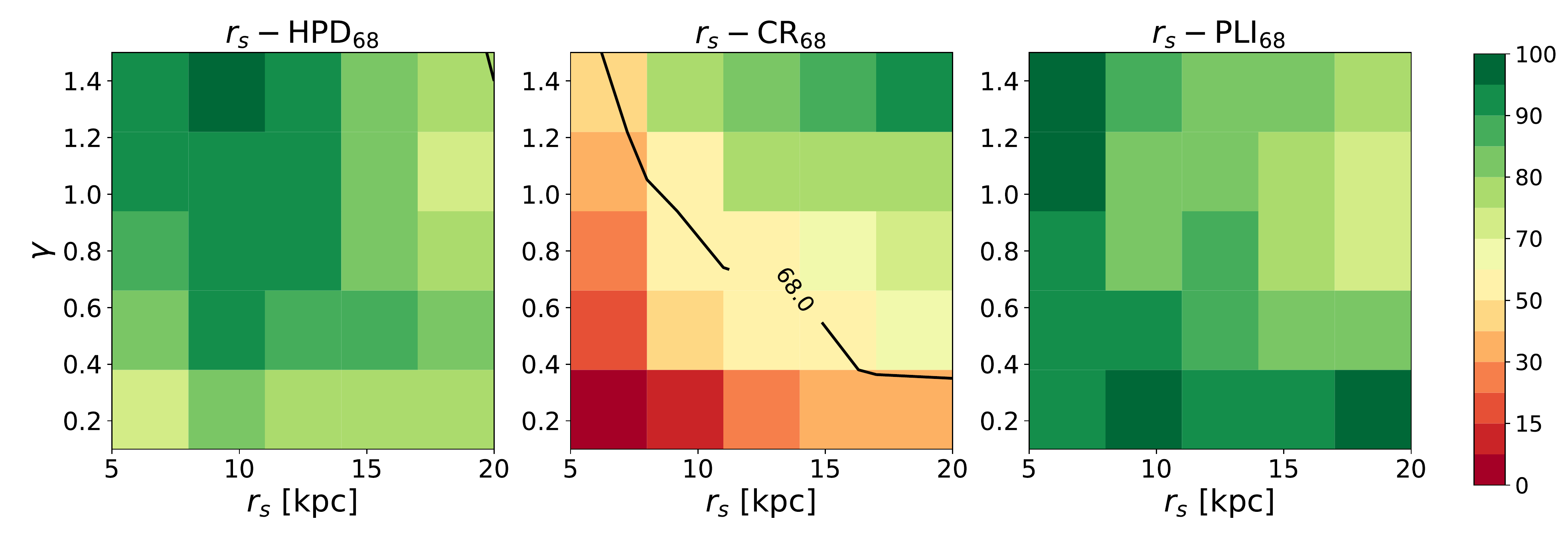}.pdf}
\caption{Coverage of intervals for the local dark matter density $\rho_0$, inner slope of the density profile $\gamma$, and the scale radius of the dark matter halo $r_s$ (top to bottom) as a function of the true values for $r_s$ and $\gamma$. From left to right, we show coverage for the 68\% highest posterior density interval (HPD), the posterior credible interval (CR) and the profile likelihood confidence interval (PL). Perfect coverage would lie along the solid black line. Green regions overcover (i.e. the intervals are too large), while orange/red regions undercover (i.e. the intervals are too short).}
\label{fig:HPD}	 
\end{centering}
\end{figure}

On Fig.~\ref{fig:HPD} we demonstrate the coverage plots from 100 mock data realization as function of the assumed fiducial parameter values, where the three columns correspond to the three interval definitions, and rows correspond to three different parameters of interest. 
In these plots, exact coverage would appear as a yellow square, with overcoverage shown in shades of green and undercoverage in orange/red. The color bars next to the plots indicate the fraction of the time, over 100 data realizations, that the inferred interval of interest contains  the true value of the parameter. The results of this figure are also summarized in Section~\ref{sec:mock data}.

\section{Dark matter distribution adopting different baryonic morphologies}
\label{App:morphologies}

Here we present the results of the MCMC analysis for all combinations of 5 disk morphologies and 6  bulge morphologies. 

In the first column of Table~\ref{tab:chi2_ML_allmorph} we list the bulge/disk morphology combinations and references on it in the literature. The details on these morphologies can be found in \cite{Iocco:2015xga,2015JCAP...12..001P}. The main conclusions here are very same as summarized in Section~\ref{sec:results} for three different disk morppholgoes, but generalized to all the possible disk and bulge morphologies and its combinations:

\begin{itemize}
    \item the value of the local density $\rho_0$ is independent of the assumed baryonic morphology
    \item the values of the reduced $\chi^2$ for all morphologies do not rule out any of them.
\end{itemize}

\begin{table}
\begin{tabular}{ | c | c | c | c | c | c | c | c |}
\hline
 Baryonic morph.    & ML $r_s$  & ML $\gamma$ & ML $\rho_0$ & HPD$_{68}$ $r_s$  & HPD$_{68}$ $\gamma$ & HPD$_{68}$ $\rho_0$ & $\chi^2_\mathrm{red}$ \\
\hline 
\hline
G2~\cite{Stanek:1996qy}BR~\cite{2013ApJ...779..115B}&6.5 & 0.01  & 0.43 & $7.4^{+1.9}_{-1.4}$ & <0.91 (95\%) & $0.43^{+0.02}_{-0.02}$ & 0.94 \\
\hline
E2~\cite{Stanek:1996qy}BR~\cite{2013ApJ...779..115B}&6.4 & 0.01  & 0.43 & $7.1^{+2.3}_{-1.1}$ & <0.93 (95\%) & $0.42^{+0.02}_{-0.02}$ & 0.91 \\
\hline
V~\cite{Vanhollebeke}BR~\cite{2013ApJ...779..115B}&6.4 & 0.0  & 0.43 & $7.3^{+2.2}_{-1.3}$ & <0.99 (95\%) & $0.43^{+0.02}_{-0.02}$ & 0.94 \\
\hline
BG~\cite{Bissantz:2001wx}BR~\cite{2013ApJ...779..115B}&6.5 & 0.0  & 0.43 & $7.1^{+2.0}_{-1.0}$ & <0.86 (95\%) & $0.42^{+0.02}_{-0.02}$ & 0.94 \\
\hline
Z~\cite{Zhao:1995qh}BR~\cite{2013ApJ...779..115B}&6.4 & 0.01  & 0.43 & $7.4^{+1.9}_{-1.5}$ & <0.93 (95\%) & $0.43^{+0.02}_{-0.02}$ & 0.94 \\
\hline
R~\cite{Robin}BR~\cite{2013ApJ...779..115B}&6.3 & 0.01  & 0.43 & $7.1^{+2.2}_{-1.0}$ & <0.91 (95\%) & $0.43^{+0.02}_{-0.02}$ & 0.94 \\
\hline
G2~\cite{Stanek:1996qy}HG~\cite{Han:2003ws}&10.4 & 1.0  & 0.44 & $9.3^{+6.8}_{-4.0}$ & $1.44^{+0.11}_{-0.54}$ & $0.43^{+0.02}_{-0.02}$ & 0.96 \\
\hline
E2~\cite{Stanek:1996qy}HG~\cite{Han:2003ws}&8.7 & 0.80  & 0.44 & $6.3^{+7.2}_{-1.4}$ & $1.20^{+0.25}_{-0.58}$& $0.44^{+0.02}_{-0.03}$ & 0.94 \\
\hline
V~\cite{Vanhollebeke}HG~\cite{Han:2003ws}&10.6 & 1.05  & 0.43 & $9.6^{+6.8}_{-3.8}$ & $1.39^{+0.17}_{-0.46}$ & $0.43^{+0.02}_{-0.02}$ & 0.95 \\
\hline
BG~\cite{Bissantz:2001wx}HG~\cite{Han:2003ws}&10.0 & 0.96  & 0.43 & $9.2^{+6.1}_{-4.3}$ & $1.28^{+0.23}_{-0.47}$ & $0.42^{+0.03}_{-0.02}$ & 0.95 \\
\hline
Z~\cite{Zhao:1995qh}HG~\cite{Han:2003ws}&10.5 & 1.03  & 0.44 & $9.6^{+5.8}_{-4.6}$ & $1.25^{+0.28}_{-0.42}$ & $0.43^{+0.02}_{-0.02}$ & 0.94 \\
\hline
R~\cite{Robin}HG~\cite{Han:2003ws}&9.9 & 0.96  & 0.44 & $8.7^{+5.6}_{-3.7}$ & $1.25^{+0.23}_{-0.51}$ & $0.44^{+0.02}_{-0.03}$ & 0.94 \\
\hline
G2~\cite{Stanek:1996qy}CM~\cite{2011MNRAS.416.1292C}&14.3 & 1.39  & 0.43 & $8.2^{+8.9}_{-2.1}$ & $1.46^{+0.16}_{-0.33}$ & $0.43^{+0.02}_{-0.02}$ & 1.00 \\
\hline
E2~\cite{Stanek:1996qy}CM~\cite{2011MNRAS.416.1292C}&10.1 & 1.11  & 0.44 & $7.9^{+6.9}_{-3.0}$ & $1.44^{+0.16}_{-0.48}$ & $0.44^{+0.02}_{-0.02}$ & 1.01 \\
\hline
V~\cite{Vanhollebeke}CM~\cite{2011MNRAS.416.1292C}& 14.2 & 1.38  & 0.43 & $11.5^{+8.7}_{-4.2}$ & $1.50^{+0.16}_{-0.25}$ & $0.43^{+0.02}_{-0.02}$ & 1.00 \\
\hline
BG~\cite{Bissantz:2001wx}CM~\cite{2011MNRAS.416.1292C}&11.0 & 1.19  & 0.44 & $9.3^{+6.0}_{-4.5}$ & $1.29^{+0.30}_{-0.28}$ & $0.44^{+0.02}_{-0.02}$ & 1.00 \\
\hline
Z~\cite{Zhao:1995qh}CM~\cite{2011MNRAS.416.1292C}&11.6 & 1.25  & 0.44 & $10.7^{+6.7}_{-5.6}$ & $1.51^{+0.12}_{-0.41}$ & $0.44^{+0.01}_{-0.03}$ & 1.00 \\
\hline
R~\cite{Robin}CM~\cite{2011MNRAS.416.1292C}&11.3 & 1.20  & 0.44 & $10.7^{+4.6}_{-5.7}$ & $1.35^{+0.24}_{-0.35}$ & $0.43^{+0.02}_{-0.02}$ & 1.00 \\
\hline
G2~\cite{Stanek:1996qy}dJ~\cite{2010ApJ...714..663D}&13.1 & 1.29  & 0.43 & $10.2^{+9.6}_{-3.8}$ & $1.47^{+0.19}_{-0.32}$ & $0.43^{+0.03}_{-0.02}$ & 0.99 \\
\hline
E2~\cite{Stanek:1996qy}dJ~\cite{2010ApJ...714..663D}&9.9 & 1.07  & 0.44 & $7.8^{+7.3}_{-2.5}$ & $1.32^{+0.23}_{-0.41}$ & $0.43^{+0.02}_{-0.02}$ & 0.98 \\
\hline
V~\cite{Vanhollebeke}J~\cite{2010ApJ...714..663D}&14.7 & 1.38  & 0.43 & $11.8^{+6.6}_{-5.3}$ & $1.45^{+0.17}_{-0.33}$ & $0.43^{+0.02}_{-0.03}$ & 0.99 \\
\hline
BG~\cite{Bissantz:2001wx}dJ~\cite{2010ApJ...714..663D}&11.5 & 1.20  & 0.43 & $10.3^{+6.4}_{-5.0}$ & $1.41^{+0.21}_{-0.36}$ & $0.43^{+0.02}_{-0.02}$ & 0.99 \\
\hline
Z~\cite{Zhao:1995qh}dJ~\cite{2010ApJ...714..663D}&12.3 & 1.24  & 0.43 & $11.1^{+6.9}_{-4.8}$ & $1.48^{+0.13}_{-0.38}$ & $0.43^{+0.02}_{-0.02}$ & 0.99 \\
\hline
R~\cite{Robin}dJ~\cite{2010ApJ...714..663D}&11.4 & 1.17  & 0.44 & $8.6^{+7.2}_{-3.8}$ & $1.31^{+0.29}_{-0.33}$ & $0.44^{+0.02}_{-0.02}$ & 0.99 \\
\hline
G2~\cite{Stanek:1996qy}J~\cite{Juric:2005zr}&15.2 & 1.31  & 0.42 & $11.1^{+8.0}_{-4.7}$ & $1.41^{+0.18}_{-0.33}$ & $0.42^{+0.02}_{-0.02}$ & 0.96 \\
\hline
E2~\cite{Stanek:1996qy}J~\cite{Juric:2005zr}&11.0 & 1.05  & 0.43 & $7.8^{+8.0}_{-2.6}$ & $1.36^{+0.20}_{-0.55}$ & $0.43^{+0.02}_{-0.03}$ & 0.95 \\
\hline
V~\cite{Vanhollebeke}J~\cite{Juric:2005zr}&14.8 & 1.31  & 0.42 & $12.5^{+6.9}_{-6.1}$ & $1.39^{+0.20}_{-0.32}$ & $0.42^{+0.02}_{-0.02}$ & 0.96 \\
\hline
BG~\cite{Bissantz:2001wx}J~\cite{Juric:2005zr}&12.3 & 1.14  & 0.42 & $10.2^{+6.7}_{-4.8}$ & $1.39^{+0.18}_{-0.42}$ & $0.42^{+0.02}_{-0.02}$ & 0.96 \\
\hline
Z~\cite{Zhao:1995qh}J~\cite{Juric:2005zr}&13.3 & 1.23  & 0.42 & $9.4^{+7.2}_{-3.7}$ & $1.48^{+0.11}_{-0.51}$ & $0.42^{+0.03}_{-0.02}$ & 0.96 \\
\hline
R~\cite{Robin}J~\cite{Juric:2005zr}&11.9 & 1.14  & 0.43 & $8.1^{+8.9}_{-1.8}$ & $1.32^{+0.21}_{-0.43}$ & $0.43^{+0.02}_{-0.02}$ & 0.96 \\
\hline

\end{tabular}
\caption{\label{tab:chi2_ML_allmorph} The maximum likelihood (ML) and the maximum a posteriori (MAP) estimates with uncertainties obtained from the 68\% HPD region (except for $\gamma$ in the BR disk morphology cases, where we indicate the 95\% upper limit as the 1D marginal posterior is compatible with~0). We also give the reduced $\chi^2$ values for the best fit (i.e. ML) parameters in the last column.}
\end{table}

\clearpage
\bibliographystyle{JHEP} 
\bibliography{biblio} 

\end{document}